\documentclass[aps,pre,groupedaddress,notitlepage,10pt]{revtex4-2}
\usepackage[utf8]{inputenc}
\usepackage{subfig}
\RequirePackage{fix-cm}
\usepackage[table]{xcolor}
\usepackage{graphicx}
\usepackage{braket}
\usepackage{qcircuit}
\usepackage{csquotes}
\usepackage{float}
\usepackage{ragged2e}
\usepackage{multirow}
\usepackage{array}
\usepackage{dcolumn}
\usepackage[colorlinks=true, citecolor=blue]{hyperref}
\usepackage{hhline}
\usepackage{amssymb}
\usepackage{amsmath}
\usepackage{collectbox}
\usepackage[utf8]{inputenc}
\usepackage{tikz}
\usetikzlibrary{shapes,arrows,positioning}
\begin{document}
	\large
	\title{Spins-based Quantum Otto Engines and Majorisation}
	\author{Sachin Sonkar} 
	\email[e-mail: ]{sachisonkar@gmail.com}
	\affiliation{ Department of Physical Sciences, 
		Indian Institute of Science Education and Research Mohali,  
		Sector 81, S.A.S. Nagar, Manauli PO 140306, Punjab, India} 
	\author{Ramandeep S. Johal} 
	\email[e-mail: ]{rsjohal@iisermohali.ac.in}
	\affiliation{ Department of Physical Sciences, 
		Indian Institute of Science Education and Research Mohali,  
		Sector 81, S.A.S. Nagar, Manauli PO 140306, Punjab, India}
	\begin{abstract} 
		The concept of majorisation is explored as a tool
		to characterize the performance of
		a quantum Otto engine in the quasi-static regime.  
		For a working substance in the form of
		a spin of arbitrary magnitude, majorisation yields a 
		necessary and sufficient condition for 
		the operation of the Otto engine, provided  
		the canonical distribution of the working medium  
		at the hot reservoir is majorised by its  canonical
		distribution at the cold reservoir.  
		For the case of a spin-1/2 
		interacting with an arbitrary spin via isotropic
		Heisenberg exchange interaction, we derive sufficient criteria for positive work extraction using the majorisation relation.
		Finally, local thermodynamics of spins as well as an upper bound on the quantum 
		Otto efficiency is analyzed using the majorisation relation.
	\end{abstract}	
	\maketitle
	\section{Introduction}
	The fast-growing field of quantum thermodynamics brings together 
	methods and tools from a variety of research areas ranging 
	from quantum information,
	open quantum systems, quantum optics, non equilibrium thermodynamics,
	theory of fluctuations, estimation theory 
	and so on \cite{Vinjanampathy_2016,Kosloff_2013,Millen_2016,quan2007quantum,Threelevelmaser,alicki2014quantum,rivas2020strong,goold2016role,PhysRevE.103.032130,Albash_2012,RevModPhys.91.025001,Esposito_2010,RevModPhys.81.1,Parrondo,Popescu2006,Alicki1979TheQO, AJM2008,PRXQuantum.3.030315,RevModPhys.93.035008,esposito2015quantum, VTJ2020,  Maffei2021, Rubio2021, LandiPRA2022}. The quantum extensions of the 
	concepts of heat, work, and entropy have in turn led to generalizations
	of the classical heat cycles. The so-called 
	quantum thermal machines exploit new thermodynamic
	resources such as quantum entanglement, coherence,
	quantum interactions and quantum statistics \cite{PhysRevA.99.062103,Alet2021,zhang2007four,albayrak2013entangled,Diaz_de_la_Cruz_2014,Pekola2015, Myers_BOF2020, 
	WatanabePRL2020}. For instance, 
	quantum Otto engine (QOE) based on various platforms has
	been widely studied for its possible quantum 
	advantages---both in its quasi-static formulations
	as well as the ones based on time-dependent constraints
	\cite{PhysRevE.90.062134,PhysRevE.101.012116,das2020quantum,chand2021finite,
		lee2020finite,turkpencce2019coupled,thomas2014friction,
    geva1992quantum,feldmann2000performance,ccakmak2017irreversible,Shastri2022,PhysRevB.101.054513,Papadatos_2021,e21111131}. 
	A QOE offers conceptual simplicity by virtue of 
	 a clear separation of heat and work steps in its heat cycle. 
	The quantum working medium used in these models may be
	taken in the form of spins, quantum harmonic oscillator, 
	interacting systems and so on \cite{e22070755,lin2003performance,Rezek2006,
		zhang2008entangled,hubner2014spin,
		azimi2014quantum,insinga2016thermodynamical,mehta2017quantum}. 
	Further, theoretical models have
	motivated experimental realizations which promise
	to be a boost for future applications in devices \cite{peterson2019experimental, MyersAVS2022}. 
	
	One of the prominent analytical tools
	guiding the theoretical developments is 
	the notion of majorisation \cite{Marshallmajorisationbook,TSagava,
		Bhatia1996MatrixA,PhysRevA.95.012110,Joe1990MajorisationAD,
		2020,vedral2015,10.5555/2011326.2011331} and its generalizations \cite{PhysRevResearch.2.033455,2016,RUCH1980222}. 
	The concept, originally perhaps from matrix analysis, 
	finds  wide applications in various areas of science, mathematics, economics, social sciences, including modern applications to  
	entanglement theory and thermodynamic resource theories.
	The majorisation partial order was developed to quantify the notion
	of disorder, in a relative sense, when comparing probability distributions. Transformations between pure 
	bipartite states by means of local operations and classical 
	communication can be determined in terms of majorisation of the 
	Schmidt coefficients of the states \cite{PhysRevLett.83.436,
		PhysRevA.91.052120,PhysRevLett.83.3566,Horodecki2003}. Majorisation has been 
	shown to determine
	the possibility of state transformations in the resource theories 
	of entanglement, coherence and purity.
	It also provides the first 
	complete set of necessary and sufficient conditions for arbitrary 
	quantum state transformations under thermodynamic processes
	\cite{majorisationcomplete,2015,2013}, which rigorously accounts for quantum 
	coherence among other quantum mechanical properties. 
	
	A majorisation relation may be defined in one of the equivalent ways, as follows.
	Suppose $x^{\downarrow}=(x_{1},...,x_{n}$) and $y^{\downarrow}=(y_{1},...,y_{n}$) are two real $n$-dimensional vectors, where $x_{}^{\downarrow}$ and $ y_{}^{\downarrow}$ indicate 
	that the elements are taken in the descending order.
	Then, the vector $x$ is said to be majorised by the vector $y$, 
	denoted as $x\prec y$, if,
	for each $m = 1,\dots, n$, we have
	\begin{equation}
		\sum_{k=m}^{n}x_{k}^{} \geq \sum_{k=m}^{n}y_{k}^{}, 
	\end{equation}
	with equality holding for $m=1$. The notion of majorisation can
	be readily applied to compare how two probability distributions0 deviate
	from a uniform distribution $u = (1/n,\dots,1/n)$. 
	Thus, $x\prec y$ implies that the distribution $y$ is more ordered
	than $x$. An important consequence of this relation is 
	the inequality: $\sum_{k=1}^{n} f(x_k) \geq \sum_{k=1}^{n} f(y_k)$, where  
	$f$ is any continuous, real-valued concave function. 
	For example, the relation $x\prec y$ implies that 
	the corresponding
	Shannon entropies are related as: $S(x) \geq S(y)$, where 
	$S(x) = - \sum_{k=1}^{n} x_k \ln x_k$. 
	Further, there are many equivalent characterizations of the majorisation 
	relation. For instance, $x$ is majorised by $y$ only when $y$ 
	can be obtained from $x$ by the action of a bistochastic matrix
	\cite{Marshallmajorisationbook}.
	
	In the present work, we characterize the 
	operation of a QOE through the notion of majorisation. 
	We show that a spin-based quantum working substance
	provides a natural platform by which the majorisation conditions 
	characterize the operation of
	a QOE. Thus, majorisation provides  
	sufficient criteria for the operation of a spin-based Otto engine.
	In fact, the analysis can be extended to a model of two spins 
	coupled via Heisenberg exchange interactions. Further, majorisation 
	provides insight into the local thermodynamics of individual
	spins in the coupled model. Using the majorisation conditions,
	we also validate an upper bound for Otto efficiency in 
	the coupled case, which is 
	tighter than the Carnot value. 
	
	The paper is organized as follows. In Section II, 
	we describe the quantum Otto cycle and its various
	stages based on a quantum 
	working substance.
	In Section III, we express the work output in terms 
	of the relative entropy between the two  equilibrium 
	distributions corresponding to hot and cold reservoirs,
	and we show that a greater value of the
	Shannon entropy of the system at the hot reservoir
	(as compared to the cold reservoir) does not ensure that
	net work may be extracted in the Otto cycle. 
	In Section IV, we show how the majorisation 
	relation, $P \prec P'$, between the hot and cold
	reservoir equilibrium distributions lead to 
	positive work condition for the QOE. 
	This is shown in Section IV.A for a single spin-$s$.
	In Section V, we analyze the coupled spins model,
	showing our main results for a special case of $(1/2,1)$
	system. In Section VI, local work by individual
	spins is analyzed based on global conditions. 
	Lastly, Section VII shows proves the enhancement
	in Otto efficiency based on majorisation conditions.
	We end our paper by summarizing our main results
	in Section VIII. The derivations of various results are
	presented in Appendix.	
	\section{Quantum Otto Engine (QOE)}
	The classical Otto cycle is a textbook example of
	a four-step heat cycle in which a classical 
	working medium, in the form of a gas or air-fuel mixture, 
	undergoes two adiabatic and two isochoric steps \cite{Zemansky}. 
	A QOE is based on  
	a quantum generalization of the classical cycle,
	in which the quantum working substance 
	(referred to below as the system)
	undergoes two quantum adiabatic steps and two isochoric
	steps. Here, we are interested in a quasi-static Otto cycle 
	in which each of the steps can take an arbitrarily long time.
	A quantum adiabatic process, either in the  
	compression or expansion stage to be explained
	below, is performed by
	varying an externally controllable parameter. 
	Secondly, for such a process, 
	the quantum adiabatic theorem \cite{Adiabatic_Fock} is assumed to hold so 
	the process does not cause any transitions between
	the energy levels, thus preserving their occupation probabilities.
	The remaining two steps are 
	the isochoric heating and cooling processes which involve thermal
	interaction of the system with the hot or cold reservoir. 
	Here, the system gets enough time to reach thermal 
	equilibrium with the corresponding reservoir. 
	The heat cycle is described in a more quantitative 
	detail as follows.

	\begin{figure}[h]
		\centering
		{\includegraphics[width=0.45\linewidth]{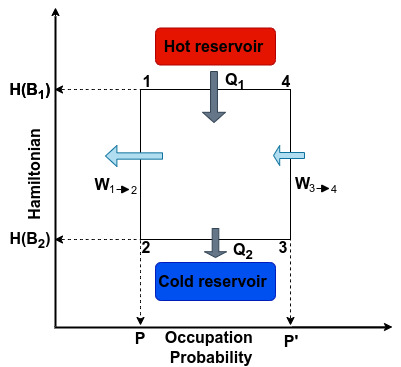}\label{fig:fig1}}
		\caption{Schematic of a quantum Otto cycle consisting of the stages 
			$1\to 2\to 3 \to 4 \to 1$.
			Steps $1\to 2$ 
			and $3\to 4$ respectively denote quantum adiabatic expansion
			and compression processes, which preserve occupation probability  and involve work $W_{1\to 2}^{} <0$
			and $W_{3\to 4}^{} > 0$. 
			The net work extracted in one cycle is $W = W_{1\to 2}^{} + W_{3\to 4}^{} < 0$. In step $4 \to 1$, heat 
			$Q_1 >0$ enters the working substance from a hot reservoir,
			while in step $2 \to 3$, heat 
			$Q_2 < 0$ is rejected to the cold reservoir. 
			Conservation of energy implies, 
			$Q_1 + Q_2 + W =0$.
		}
		\label{fig:fig3}
	\end{figure}
	\par\noindent
	\textbf{Stage 1.}
	Consider an $n$-level quantum system
	with Hamiltonian $H(B_{1})$ whose eigenvalues  
	can be arranged (in descending order) as:
	$\varepsilon^{\downarrow} = 
	(\varepsilon_{n}, \dots, \varepsilon_{1})$.
	The system is in thermal equilibrium 
	with the hot reservoir at 
	temperature $T_1$. The canonical occupation probabilities 
	for different energy levels, $P_{k}=e^{-\varepsilon_{k}/T_{1}}/{\sum_{k} e^{-\varepsilon_{k}/T_{1}}}$, are arranged as 
	$P^{\downarrow}=(P_{1},\dots,P_{n}$),
	where we have set the Boltzmann constant $k_{\rm B}$ equal to unity.
	The energy eigenstates are represented by 
	$\{\ket{\psi_{k}} \vert k=1,...n\} $. 
	Thus, the density matrix representing the thermal 
	state of the system is given by:
	\begin{equation} \label{eqn6}
		\rho=\sum_{k=1}^{n}P_{k}\ket{\psi_{k}}\bra{\psi_{k}}, 
	\end{equation}
	\textbf{Stage 2.} The system is detached from the 
	hot reservoir and undergoes a quantum adiabatic process
	in which the external field strength is {\it lowered} from $B_{1}$ to 
	$B_{2}$. Here, the quantum adiabatic theorem 
	ensures that no transitions are induced between
	the energy levels in the change from  
	$\varepsilon_{k}$ to $\varepsilon_{k}^{\prime}$. Suppose that the energies after the first adiabatic process
	are given by:
	${\varepsilon^{\prime}}^{\downarrow} = 
	(\varepsilon_{n}^{\prime}, \dots, \varepsilon_{1}^{\prime})$,
	where we assume no level-crossing as the Hamiltonian changes from $H(B_1)$ to $H(B_2)$. 
	\par\noindent
	\textbf{Stage 3.} The system is brought in thermal contact with the cold reservoir at temperature $T_{2} (<T_{1})$. The energy eigenvalues remain at $\varepsilon_{k}^{\prime}$ 
	while the occupation probabilities change from $P_{k}$ to $P_{k}^{\prime}=e^{-\varepsilon_{k}^{\prime}/T_{2}}
	/{\sum_{k} e^{-\varepsilon_{k}^{\prime}/T_{2}}}$, which are ordered as:  
	$P^{{\prime}{\downarrow}} = (P_{1}^{\prime},\dots, P_{n}^{\prime}$). 
	Thus, the density matrix of the system at the end of Stage-3
	is given by:
	\begin{equation} \label{eqn66}
		\rho^{\prime}=\sum_{k=1}^{n}P_{k}^{\prime}\ket{\psi_{k}}\bra{\psi_{k}}.
	\end{equation}
	\par\noindent
	\textbf{Stage 4.} The system is detached from the cold reservoir and the field strength is changed back 
	to $B_{1}$.
	The occupation
	probabilities $\{P_{k}^{\prime}\}$ remain unchanged, while the energy levels change back 
	from $\{\varepsilon_{k}^{\prime}\}$ to
	$\{\varepsilon_{k}\}$. 
	
	Finally, the system is attached to the hot reservoir again whereby 
	the initial state ($\rho$) is recovered, thus 
	completing one heat  cycle. Note that only heat is exchanged between the system and the reservoir 
	during an isochoric process, 
	which is given by  the difference between the final and initial
	mean energies of the system in that process. Thus, in
	Stage-1 and Stage-3, the heat exchanged is given respectively as:
	\begin{align} \label{eqn7}
		Q_{1}&=\sum_{k=1}^n\varepsilon_{k}(P_{k}-P_{k}^{\prime}),~~~~~
		Q_{2} =\sum_{k=1}^n\varepsilon_{k}^{\prime}(P_{k}^{\prime}-P_{k}).
	\end{align}
	On the other hand, only work is performed during 
	the adiabatic branches of the
	quantum Otto cycle. Let $W$ be the net work performed in one cycle. Applying 
	the law of conservation of energy to the cyclic process, 
	we have: $Q_1 + Q_2 + W =0$.
	The operation of a heat engine requires that 
	heat is absorbed (rejected) 
	by the system at the hot (cold) reservoir,
	while net work is extracted from the system by 
	the end of the cycle.
	These conditions can be satisfied by choosing the sign convention: $Q_1 >0$, $Q_2 <0$
	and $W <0$. The net
	work performed by the QOE can then be written as
	\begin{equation} \label{eqn8}
		\vert W \vert = Q_{1}+Q_{2} = \sum_{k=1}^{n}(\varepsilon_{k}-
		\varepsilon_{k}^{\prime})(P_{k}-P_{k}^{\prime}).
	\end{equation}
	We denote $\vert W \vert \geq 0$ as the positive work condition (PWC) 
	of our engine. The efficiency of the QOE
	is defined as $\eta=\vert W \vert/Q_{1} = 1+ Q_2/Q_1$.
	
	\section{Relative entropy and QOE}
	In this section, we cast the thermodynamic
	quantities for a QOE in terms of the relative 
	entropy which is defined as
	$D(x||y) \equiv \sum_k x_{k} \left(
	\ln x_{k} - \ln y_{k} \right) \geq 0$.
	Also known as the Kullback-Leibler divergence\cite{10.1214/aoms/1177729694},
	this quantity is a measure of the 'distance'
	between two discrete probability distributions, 
	and vanishes only when 
	the distributions $x$ and $y$ are identical.
	
	Now, the expressions for canonical probabilities 
	may be inverted as: 
	$\varepsilon_{k} \equiv -T_{1}\ln{(P_{k} 
		\sum_{j} e^{-\varepsilon_{j}/T_{1}})}$ and 
	$\varepsilon_{k}^{\prime} \equiv -T_{2}\ln{(P_{k}^{\prime}
		\sum_{j} e^{-\varepsilon_{j}^{\prime}/T_{2}})}$.
	Substituting these expressions in 
	Eq. (\ref{eqn7}),  and after some algebra
	(Appendix A), the heat exchanged with each
	the reservoir is expressed as:
	\begin{align}
		Q_1 & = T_1 (S_1-S_2) - T_1 D(P^{\prime}||P),
		\label{q1D}
		\\
		Q_2 & =  -T_2 (S_1-S_2) - T_2 D(P||P^{\prime}),
		\label{q2D}
	\end{align}
	where $S_1 =-\sum_{k} P_{k} \ln P_{k}$ and $S_2 =-\sum_{k} P_{k}' \ln P_{k}'$
	are the Shannon entropies of the system in equilibrium
	with hot and cold reservoirs, respectively. as we have let $k_B=1$ Shannon entropy is equal to canonical entropy.
	
	Thus, it can be seen that $D(P^{\prime}||P)$
	is equal to the entropy generated in the 
	hot isochoric step. 
	$D(P||P^{\prime})$ has a similar meaning
	for the cold isochoric step.
	The net work extracted in an Otto cycle is given by:
	\begin{equation}\label{mj6}
		|W|=(T_{1}-T_{2}) (S_{1}-S_{2})-T_{1}D(P^{\prime}||P)-T_{2}D(P||P^{\prime}).
	\end{equation}		
	Using, Eq. (\ref{q1D}) and Eq. (\ref{q2D}) the total entropy generated
	in the heat cycle, 
	$\Delta_{\rm tot} S = -Q_2/T_2 - Q_1/T_1$,
	can be expressed as:
	\begin{equation}
		\Delta_{\rm tot} S = D(P||P^{\prime}) + D(P^{\prime}||P).
		\label{dstot}
	\end{equation}
	The total entropy generated in a quantum Otto cycle
	is thus equal to the symmetric sum of the relative
	entropies. This quantity is also known as the 
	symmetrized divergence and is distinguished by
	the fact that it serves as a metric in the 
	space of probability distributions. 
	Finally, the positivity of the total entropy
	generated proves the consistency of the QOE 
	with the second law of thermodynamics, and hence
	its efficiency is bounded by the Carnot value:
	$\eta \leq 1-T_2/T_1$.
	
	Returning to Eq. (\ref{mj6}), the positivity
	of the relative entropy indicates that
	for $T_1 >T_2$, 
	$S_1 >S_2$ is a necessary,  
	but not a sufficient condition for
	$|W| \geq 0$. The necessity of the condition $S_1>S_2$ 
	can be reasoned due to the fact that 
	heat is absorbed by the system at the hot reservoir,
	while heat is rejected by the system at the cold reservoir, 
	and  the intermediate, quantum adiabatic processes do not
	alter the entropy of the system.
	These considerations suggest that more general conditions 
	are desirable to characterize 
	the probability distributions, which not only 
	ensure $S_1>S_2$, but also the PWC or $|W| \geq 0$. 
	In this paper, we show
	that the majorisation relation ($P \prec  P'$) provides 
	sufficient conditions for the operation of
	a spins-based quantum Otto cycle as a heat engine.
	
	\section{Majorisation and QOE}
	As mentioned earlier, the majorisation relation $P\prec P^{\prime}$
	implies the following set of inequalities:
	\begin{equation}
		\sum_{k=m}^{n}P_{k}^{} \geq \sum_{k=m}^{n}P_{k}^{\prime},
		\quad (m =1,\dots,n)
		\label{majn2}
	\end{equation}
	with the equality holding for $m=1$ owing to the normalization
	property of each distribution.
	Specifically, we obtain from the above inequalities, for
	$m=n$
	\begin{equation} \label{eqna}
		P_{n}\geq P_{n}^{\prime},
	\end{equation} 
	and, for $m=2$, along with normalization 
	\begin{equation} \label{majn1}
		P_{1}^{\prime} \geq P_1.
	\end{equation} 
	These inequalities may be combined as:
	$ {P_{1}^{\prime}}/{P_{n}^{\prime}} \geq {P_{1}}/{P_{n}}$, 
	to yield the condition:
	\begin{equation} \label{majcond1}
		\frac{\varepsilon_{n}^{\prime}- \varepsilon_{1}^{\prime}}{T_{2}}
		\geq \frac{\varepsilon_{n}^{}- \varepsilon_{1}^{}}{T_{1}}.
	\end{equation}
	For $T_1>T_2$, the above inequality will yield a nontrivial condition, 
	provided that $\varepsilon_{n} - \varepsilon_{1}>\varepsilon_{n}^{\prime}- 
	\varepsilon_{1}^{\prime}$. In other words, we must assume that the range of 
	the energy spectrum shrinks during the first quantum adiabatic process. Apart from that, the condition (\ref{majcond1}) 
	is derived for a generic, non-degenerate spectrum. 
	
	Now, an important question arises regarding the circumstances 
	under which the majorisation inequalities, Eq. (\ref{majn2}), hold.
	Naturally, this is dependent on the form of Hamiltonian
	(or the energy spectrum which enters the expressions for
	the canonical probabilities). 
	In the following, we show for a spin system, 
	a set of necessary
	and sufficient conditions to satisfy the majorisation relation.

	\subsection{QOE with a single spin-$s$}
	
	Suppose the system is in the form of
	a quantum spin of magnitude $s$. The energy
	eigenvalues in Stage-1 are: $\varepsilon_k = 2(k -s -1) B_1$, where 
	$k =1, \dots, 2s+1$. Explicitly, we have
	\begin{equation*}
		\varepsilon_{1}=-2sB_{1}, \; \varepsilon_{2}=-2(s-1)B_{1},
		\dots,
		\; \varepsilon_{2s}^{}=2(s-1)B_{1}, \; \varepsilon_{2s+1}^{}=2sB_{1}.
	\end{equation*}
	After the first quantum adiabatic step,
	the energy spectrum is given by:  $\varepsilon_{k}^{\prime} = 2(k -s -1) B_2$, where $k =1, \dots, 2s+1$ and $B_2 < B_1$.
	
	Now, for this system,
	Eq. (\ref{majcond1})
	simplifies to the following condition:
	\begin{equation}
		\frac{B_{2}}{T_{2}}\geq \frac{B_{1}}{T_{1}}.
		\label{bt12}
	\end{equation}
	The above condition was first derived in Ref. \cite{kieu2004second}
	for a two-level quantum system (equivalent to $s=1/2$). 
	Here, we see
	it as a consequence of the majorisation relation 
	between the canonical distributions corresponding to
	hot and cold reservoirs. In fact, the above condition
	is necessary and sufficient to satisfy all the majorisation 
	inequalities, (\ref{majn2}) \cite{2021}. 
	
	Next, using the definitions in Eq. (\ref{eqn7}),
	the  heat exchanged between the system 
	and each reservoir is calculated to be:
	\begin{align} \label{eqn52}
		Q_{1}&=2B_{1}X, \quad Q_{2}=-2B_{2}X,
	\end{align}
	where
		$X =\sum_{k=2}^{2s+1}(k-1)\left(P_{k}-P_{k}^{\prime}\right)$. 
	Thus, the magnitude of the work performed in one cycle is:
	\begin{equation} \label{eqn53}
		\vert W \vert=2(B_{1}-B_{2})X.
	\end{equation}
	Now, assuming the relation 
	$P\prec P^{\prime}$, and for $n=2s+1$,
	we add up all the inequalities  
	(\ref{majn2}) corresponding to $m=2,\dots,n$. 
	The resulting inequality can be rewritten in the form 
	$X \geq 0$. In other words, we obtain that
	$P\prec P^{\prime}$ implies 
	$|W| \geq 0$, provided $B_1>B_2$. 
	
	Thus, we may say that for a spin-$s$ system, 
	the majorisation relation $P\prec P^{\prime}$ 
	is a sufficient and necessary condition for the 
	operation of QOE.

	As an extreme case scenario, we may have the 
	conditions:   $P_{k} \geq P_{k}^{\prime}$,
	for $k = 2, \dots, n$.
	The normalization property then ensures $P_{1} \leq P_{1}^{\prime}$.
	It is clear that the above inequalities  
	satisfy the majorisation conditions (\ref{majn2}), 
	implying that $P \prec P^{\prime}$. 
	Thus, the above 
	extreme case, applied to the case of 
	a spin, also leads to PWC (see \cite{2021} for details). 
	
	\section{QOE with two coupled spins}
	Next, we consider a system of two coupled spins, a spin-$1/2$ particle 
	interacting with an arbitrary spin-$s$, via 1-d isotropic,  Heisenberg 
	exchange interaction. The Hamiltonian of the working substance is
	\begin{equation} \label{eqn5}
		H=2B(s_{z}^{(1)}\otimes I^{(2)}+I^{(1)} \otimes s_{z}^{(2)})+
		8J(s_{x}^{(1)}\otimes s_{x}^{(2)}+s_{y}^{(1)}\otimes s_{y}^{(2)}
		+s_{z}^{(1)}\otimes s_{z}^{(2)}),
	\end{equation}
	where $J\geq 0$ is the coupling strength parameter. Here, 
	$s^{(1)} \equiv \{s_{x}^{(1)},s_{y}^{(1)},s_{z}^{(1)}\}$ and $s^{(2)}
	\equiv \{s_{x}^{(2)},s_{y}^{(2)},s_{z}^{(2)}\}$ are the spin-1/2 and 
	spin-$s$ operators, respectively, and $I$ denotes the identity operator.
	We set $\hbar=1$, 
	Bohr magneton $\mu_{b}=1$, and
	assume there is no orbital angular momentum so that the
	gyromagnetic ratio $\gamma$ is the same for both spins, $\gamma=2$.
	The total number of levels of the bipartite system
	is $n=2(2s+1)$.
	The energy spectrum is displayed in Fig. 1 of SM. Note that
	the energy eigenvalues contain a constant term $4 s J$
	which can be adjusted, for convenience,
	by an overall shift of the energy spectrum.
	Further, note that only the field parameter
	$B$ is varied cyclically while the 
	coupling parameter $J$ is held fixed. 
	
	QOE of the above kind was first studied with two spin-1/2 particles \cite{PhysRevE.83.031135} where, amongst other things, an enhancement in Otto efficiency was reported as a result of 
	coupling between the spins. 
	Further, the model was extended incorporating 
	the above Hamiltonian \cite{PhysRevE.92.022142}. As the energy spectrum 
	becomes more complex, numerical results were used to gain insights 
	into the performance of QOE. In \cite{2021}, a heuristics-based approach was used to analyze the general case of
	spin-$s_1$ coupled with 
	spin-$s_2$. Thus, sufficient criteria for work extraction 
	were inferred using the extreme case scenarios. It was also 
	argued with numerical results that majorisation leads to 
	a more robust characterization of QOE than the extreme case
	scenario. Motivated by these findings, in this paper,
	we develop a characterization of the QOE in terms 
	of the majorisation relation. 
	
	In the following, we show how the majorisation conditions 
	also lead to sufficient criteria for QOE based
	on the above coupled-spins model.  
	In order to illustrate our main results, we treat the case
	of a spin-1/2 particle coupled to a spin-1, for which $n=6$.
	The results for the more general case of $(1/2,s)$
	system are reported in SM.
	\subsection{The coupled $(1/2,1)$ system}
	By introducing a constant shift of $4sJ \equiv 4J$ 
	in the energy eigenvalues at the hot reservoir,
	these are given by:
	$\varepsilon_{1}=-3B_{1}, \; \varepsilon_{2}=-B_{1}-12J, 
	\; \varepsilon_{3}=-B_{1},
	\; \varepsilon_{4}=B_{1}-12J,\;\varepsilon_{5}=B_{1},
	\;\varepsilon_{6}=3B_{1}$.
	The corresponding eigenstates are:
	$\ket{\psi_{1}} =\ket{12^\prime}, ~\ket{\psi_{2}}
	=(\sqrt{2}\ket{02^\prime}-\ket{11^\prime})/\sqrt{3}, ~\ket{\psi_{3}} 
	=(\ket{02^\prime}+\sqrt{2}\ket{11^\prime})/\sqrt{3},~
	\ket{\psi_{4}} 
	=(\ket{01^\prime}-\sqrt{2}\ket{10^\prime})/\sqrt{3}, 	~\ket{\psi_{5}} 
	=(\sqrt{2}\ket{01^\prime}+\ket{10^\prime})/\sqrt{3}, 	~\ket{\psi_{6}}=\ket{00^\prime}$,
	where $\ket{0} \equiv (1,0)^T$ and $\ket{1} \equiv (0,1)^T$ are eigen-kets for the bare Hamiltonian of spin-1/2. 
	Similarly, $\ket{0^{\prime}}\equiv (1,0,0)^T,~\ket{1^{\prime}} \equiv (0,1,0)^T$ and $\ket{2^{\prime}} \equiv (0,0,1)^T$ are the eigen-kets
	for the bare Hamiltonian of spin-1.
	The density matrix in the initial state
	of the coupled system is: 
	$\rho=\sum_{k=1}^{6}P_{k}\ket{\psi_{k}}\bra{\psi_{k}}$.
	Similarly at the cold reservoir, the
	density matrix is 
	$\rho^{\prime}=\sum_{k=1}^{6}P_{k}^{\prime}\ket{\psi_{k}}
	\bra{\psi_{k}}$.
	
	Now, by inspection, for $B_{1}>6J$,
	the energy levels are ordered as:
	$\varepsilon^{\downarrow} = 
	(\varepsilon_{6}, \dots,
	\varepsilon_{1})$, and
	therefore, 
	$P^{\downarrow} = (P_{1},\dots, P_{6}$).
	Similarly, for $B_{2}>6J$, we have
	${\varepsilon^{\prime}}^{\downarrow} = 
	(\varepsilon_{6}^{\prime}, \dots, \varepsilon_{1}^{\prime})$,
	as well as $P^{{\prime}{\downarrow}}=(P_{1}^{\prime},\dots, P_{6}^{\prime}$).
	Then, the heat $Q_{1} = \sum_{k=1}^{6} \varepsilon_{k}(P_{k}-P_{k}^{\prime})$ can be expressed as:
	\begin{equation} \label{eqn56}
		Q_{1}=2B_{1}{\cal X}-12 J {\cal Y},
	\end{equation}
	where
	\begin{align}	
		{\cal X}&=3(P_{6}-P_{6}^{\prime})+2(P_{5}-P_{5}^{\prime})+2(P_{4}-P_{4}^{\prime}
		)+(P_{3}-P_{3}^{\prime})+(P_{2}-P_{2}^{\prime}) \label{calx},\\
		{\cal Y} &=(P_{2}-P_{2}^{\prime})+(P_{4}-P_{4}^{\prime}).
		\label{caly}
	\end{align}	
	Similarly, we can evaluate:  
	$Q_{2}=\sum_{k=1}^{6} \varepsilon_{k}'(P_{k}-P_{k}^{\prime}) = 2B_{2}{\cal X}-12 J {\cal Y}$.
	Thus, the work extracted in one cycle is
	\begin{equation} \label{eqn58}
		|W|=2(B_{1}-B_{2}){\cal X}.
	\end{equation}
	For $B_{1}>B_{2}$, PWC requires ${\cal X} \geq 0$.
	Now, under the relation $P\prec P^{\prime}$,
	the following set of inequalities must hold:
	\begin{align}
		P_{6}&\geq P_{6}^{\prime} \label{eqn60}\\
		P_{5}+P_{6}&\geq P_{6}^{\prime}+P_{5}^{\prime} \label{eqn61}\\
		P_{4}+P_{5}+P_{6}&\geq P_{6}^{\prime}+P_{5}^{\prime}+P_{4}^{\prime} \label{eqn62}\\
		P_{3}+P_{4}+P_{5}+P_{6}&\geq 
		P_{6}^{\prime}+P_{5}^{\prime}+P_{4}^{\prime}+P_{3}^{\prime} \label{eqn63}\\
		P_{2}+P_{3}+P_{4}+P_{5}+P_{6}&\geq 
		P_{6}^{\prime}+P_{5}^{\prime}+P_{4}^{\prime}+P_{3}^{\prime}+P_{2}^{\prime}. \label{eqn64}
	\end{align}
	Further, as we have seen above,  Eq. (\ref{eqn64}) implies:
	\begin{equation} \label{eqn66}
		P_{1}\leq P_{1}^{\prime}.
	\end{equation}
	Eqs. (\ref{eqn60}) and (\ref{eqn66}) can be combined
	to yield the condition $B_2/T_2 \geq B_1/T_1$, which 
	is the same condition as for a QOE based on a single spin.
	
	Then, upon adding Eqs. (\ref{eqn60}), (\ref{eqn62}) and (\ref{eqn64}),
	we obtain  
	\begin{equation}
		P_{2}+P_{3}+2P_{4}+2P_{5}+3P_{6}\geq 
		3P_{6}^{\prime}+2P_{5}^{\prime}+2P_{4}^{\prime}+P_{3}^{\prime}+P_{2}^{\prime}, 
	\end{equation}
	which can be rearranged as the inequality: ${\cal X} \geq0$.
	In this manner, we see that
	the majorisation relation ($P \prec P'$) directly  
	implies the positive work condition for the QOE based on the coupled $(1/2,1)$ system, provided $B_1 > B_2$. The proof can 
	be straightforwardly generalized to the case of a $(1/2,s)$
	system, as discussed in Section II of SM.
	
\begin{figure}[h!]
		\centering
		{\includegraphics[width=0.7\linewidth]{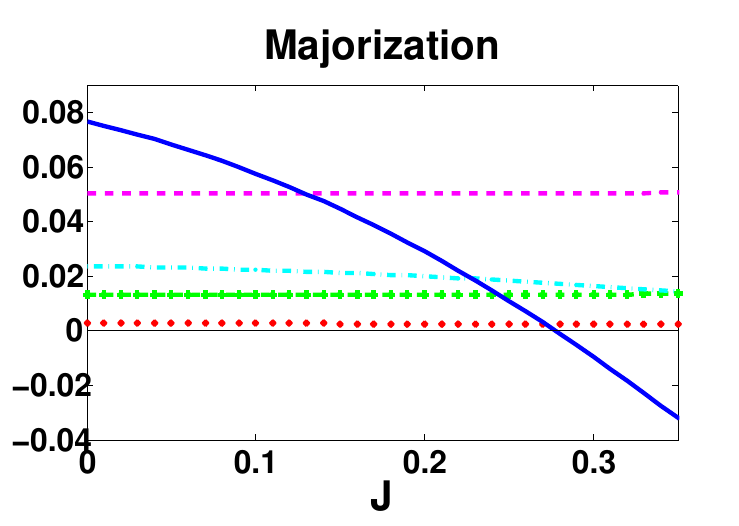}}
		\caption{{Majorisation inequalities for the coupled
				$(1/2,1)$ system, 
				(Eqs. (\ref{eqn60}) to (\ref{eqn64})). 
				$P_{6}-P_{6}^{\prime} \geq 0$ (red, dot curve), 
				$\sum_{k=5}^{6}(P_{k}-P_{k}^{\prime})\geq 0$ (green, star dashed curve), 
				$\sum_{k=4}^{6}P_{k}-P_{k}^{\prime} \geq 0$ (cyan, dot-dashed curve),
				$\sum_{k=3}^{6}P_{k}-P_{k}^{\prime}\geq 0$ (purple, dashed curve), 
				$\sum_{k=2}^{6}P_{k}-P_{k}^{\prime}\geq 0 $ (blue), 
				are together verified in a limited range of $J$ values. 
				The last inequality (blue curve) yields the 
				strictest range of $J$ values, beyond which the 
				majorisation relation ($P \prec P'$)  does not hold. 
				The parameters are chosen as $B_{1}=5,B_{2}=3,T_{1}=6,T_{2}=3$},
			such that $B_1 > B_2$ and $B_2/T_2 > B_1/T_1$.
			Analytically, we can derive a sufficient
			condition for the majorisation relation to  hold, as 
			$0\leq J \leq J_c$ where in this case $J_c=0.189$     (see Eq. (\ref{eqn69}).
		}
		\label{fig_maj21}
	\end{figure}

	Now, in order to ascertain the conditions under
	which the majorisation inequalities themselves hold,  
	for the given Hamiltonian of the system, 
	we employ numerical evidence.
	As Fig. \ref{fig_maj21} shows,
	$B_2/T_2 \geq B_1/T_1$ is a necessary, but not a sufficient
	condition in the case of the coupled system.
	The majorisation relation may be verified only 
	for a limited range of $J$ values (for given $T_1,T_2,B_1,B_2$). 
	Thus, condition (\ref{eqn64}),
	$\sum_{k=2}^{6} (P_{k}-P_{k}^{\prime})\geq 0$, 
	is violated beyond a certain range of $J$.
	Now, it is difficult to estimate this range analytically for,
	say, arbitrary reservoir temperatures. Numerically, it is seen that 
	as the temperatures of the reservoirs are raised, 
	the range of validity of the majorisation relation (in terms of 
	the range of $J$)
	also broadens. Below, we infer  a {\it sufficient} criterion for majorisation, in terms of the permissible range of $J$ values, 
	which is followed with a good accuracy at lower temperatures.
	
	Upon combining Eqs. (\ref{eqn62}) and (\ref{eqn66}), we arrive 
	at the condition:	$0 \leq J_{}\leq {\Phi}/{3}$,
	where $\Phi=({B_{2}}/{T_{2}}-{B_{1}}/{T_{1}})
	({1}/{T_{2}}-{1}/{T_{1}})^{-1}$.
	The same condition is obtained upon combining
	Eqs. (\ref{eqn63}) and (\ref{eqn66}). However, 
	the validity of Eq. (\ref{eqn66}) requires that
	\begin{equation} \label{eqn69}
		0 \leq J\leq\frac{\Phi}{6} +\frac{1}{12}\left(\frac{1}{T_2}-\frac{1}{T_1}\right)^{-1}\ln\left(\frac{1+e^{-2B_1/T_1}}{1+e^{-2B_2/T_2}}\right)\equiv J_c.
	\end{equation}
	Thus, we may infer that the above range for $J$ is the 
	strictest one for which {\it all} majorisation 
	conditions (Eqs. (\ref{eqn60}) to (\ref{eqn64}))  hold good.
	Clearly, we have $J_c \geq \Phi/6$.
	
	\begin{figure}[h!]
		\centering
		{\includegraphics[width=0.7\linewidth]{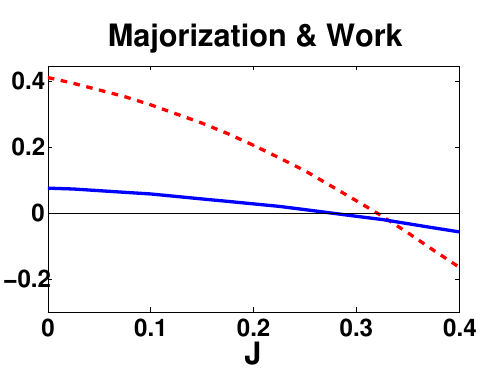}}
		\caption{Work output (Eq. (\ref{eqn58})) and the  majorisation condition
			(Eq. (\ref{eqn64})) for the coupled $(1/2,1)$ system, with the parameters set as 
			$B_{1}=5,B_{2}=3,T_{1}=6,T_{2}=3$.
			The majorisation relation (blue curve, as in Fig. 
			\ref{fig_maj21}) is violated beyond a certain 
			value of the coupling strength $J$. However, the 
			work output can be nonzero even beyond that range.} 
		\label{fig_wmj}
	\end{figure}
	In summary,
	we have shown that if the majorisation
	relation ($P \prec P'$) is satisfied, then the model works like an engine (PWC). 
	However, it is important to note that
	the PWC {\it can} hold 
	even if the majorisation relation is not satisfied,
	as depicted in Fig. \ref{fig_wmj}.
	For the coupled $(1/2,1)$
	model, it implies that net work may be obtained 
	(${\cal X} > 0$) for $J > J_c$. 
	Given that $B_1 > B_2$ along with the 
	conditions (\ref{bt12}) and (\ref{eqn69}), all majorisation inequalities hold  and thus 
	yield sufficient criteria for PWC in the case  of $(1/2,1)$ system.
	On the other hand, for the single spin, we obtained
	sufficient {\it and} necessary conditions from majorisation.
	Finally, based on induction, 
	we can infer sufficient criteria for the general case 
	of $(1/2,s)$ system, which are 
	the conditions: $B_1 > B_2$, $B_2/T_2 \geq B_1/T_1$
	and  $0 \leq J \leq {\cal J}_c$,
	within which PWC holds for the $(1/2,s)$ system.
	The details are mentioned in Appendix B.

	\section{Analysis of local work}
\begin{figure}[h!]
		\centering
		{\includegraphics[width=0.7\linewidth]{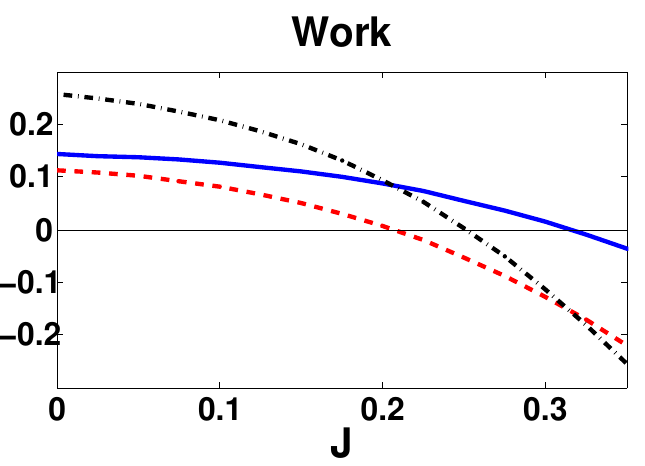}}
		\caption{ Global (black, dot-dashed curve) 
		versus the local work for spin-1/2 (red, dashed curve)
		and spin-1 (blue curve) for the
			coupled $(1/2,1)$ system with parameters set as: 
			$B_{1}=5,B_{2}=3,T_{1}=4,T_{2}=2$. Global 
			work is the sum of local contributions by
			the two spins. As $J$ value is increased, 
			spin-1/2 ceases to yield work first. 
			Spin-1 is able to extract work even at
			higher $J$ values. These features
			can be analytically inferred, as described in Section VI
			} 
		\label{fig_glw}
	\end{figure}

	Now, each spin constituting the bipartite coupled system 
	is governed by a local Hamiltonian which also depends
	on the parameter $B$, and so undergoes a cyclic evolution.
	Thus, it is of intact to examine the local performance
	of each spin in the quantum Otto cycle. The local 
	state of each spin is obtained from its
	reduced density matrix. Thus, 
	upon summing over the degrees of freedom of spin-1,
	the reduced density matrix for spin-1/2, in Stage-1, is
	defined as:
		$\rho^{(1/2)}=\sum_{m=0^\prime}^{2^\prime}\bra{I\otimes 
			m}\rho\ket{I\otimes m}$,
	which may be written in diagonal form, 
	as: $\rho^{(1/2)} = \{ q_1, q_2 \}$,
	where $q_2$ is the occupation probability
	of the excited state, given by
	\begin{align}	
		q_{2}&=\frac{2P_{2}}{3}+\frac{P_{3}}{3}+\frac{P_{4}}{3}+\frac{2P_{5}}{3}+P_{6} \label{eqn82},
	\end{align}
	and $q_1 = 1-q_2$.
	The work performed by spin-1/2 is evaluated to be:
	$W_{1/2}^{}=2(B_{1}-B_{2})(q_{2}-q_{2}^{\prime})$.
	For convenience,  we express this work in the form:	
	\begin{equation} \label{eqn91}
		W_{1/2}=\frac{2}{3} (B_{1}-B_{2})({\cal X} + {\cal Z}),
	\end{equation}
	where ${\cal X}$ has been defined in Eq. (\ref{calx}), and
	\begin{equation} \label{eqn106}
		{\cal Z}=(P_{2}-P_{2}^{\prime})-(P_{4}-P_{4}^\prime).
	\end{equation}
	Similarly, the reduced density matrix for spin-1, 
	$\rho^{(1)}=\sum_{m=0}^{1}\bra{m\otimes I}\rho\ket{m\otimes I}$,
	is written as: $\rho^{(1)} \equiv\{r_1,r_2, r_3  \}$, where the occupation probabilities:
	\begin{align}
		r_{1}&=P_{1}+\frac{2P_{2}}{3}+\frac{P_{3}}{3}, \label{eqn84}\\
		r_{2}&=\frac{P_{2}}{3}+\frac{2P_{3}}{3}+\frac{P_{4}}{3}+\frac{2P_{5}}{3}, \label{eqn85}\\
		r_{3}&= \frac{2P_{4}}{3}+\frac{P_{5}}{3}+P_{6}, \label{eqn86}
	\end{align}
	are ordered as: $r_1 > r_2 > r_3$.
	Then, the local work by spin-1 is given by:
	$W_{1}=2(B_{1}-B_{2})[2(r_{3}-r_{3}^{\prime}) + (r_{2}-r_{2}^{\prime})]$,
	which can be rewritten in the form:
	\begin{equation} \label{eqn105}
		W_{1}=\frac{2}{3} (B_{1}-B_{2})(2 {\cal X} - {\cal Z}),
	\end{equation}
	where we have used Eqs. (\ref{eqn85}) and (\ref{eqn86})
	along with the definitions of $\cal X$ and $\cal Z$.
	Now, it is easily verified that the local work contributions, 
	Eqs. (\ref{eqn91}) and (\ref{eqn105}),  
	add up to yield the global work (Eq. (\ref{eqn58})), 
	i.e. $W_{1/2} + W_1 = |W|$. In Fig. \ref{fig_glw},
	we compare the local and global work output.
	It is observed that
	the spin-1/2 work vanishes prior to the spin-1
	work. This observation may be justified as
	follows.

	As discussed earlier, 
	for $B_1 > B_2$ and $T_1 > T_2$, 
	the majorisation relation ($ P \prec P'$) suggests
	the following set of sufficient conditions:  $B_2/T_2  \geq B_1/T_1$
	and $0 \leq J \leq J_c$. From the majorisation
	inequalities, we have derived, in Section V.A, 
	the condition ${\cal X} \geq 0$ or PWC
	for the global system. 
	Furthermore, it can also be shown that sufficient conditions for PWC in spin-1/2 and spin-1 are respectively (see Appendix C.1), 
	\begin{equation} \label{eqnJmx1}
		0 \leq J\leq\frac{\Phi}{6} +\frac{1}{12}\left(\frac{1}{T_2}-\frac{1}{T_1}\right)^{-1}\ln\left(\frac{2+e^{-2B_1/T_1}}{2+e^{-2B_2/T_2}}\right)\equiv J_c^{(1/2)}.
	\end{equation}
	\begin{equation} \label{eqnJmx2}
		0 \leq J\leq\frac{\Phi}{6} +\frac{1}{12}\left(\frac{1}{T_2}-\frac{1}{T_1}\right)^{-1}\ln\left(\frac{1+5e^{-2B_1/T_1}}{1+5e^{-2B_2/T_2}}\right)\equiv J_c^{(1)}.
	\end{equation}
	
\begin{figure}[h!]
		\centering
		{\includegraphics[width=0.7\linewidth]{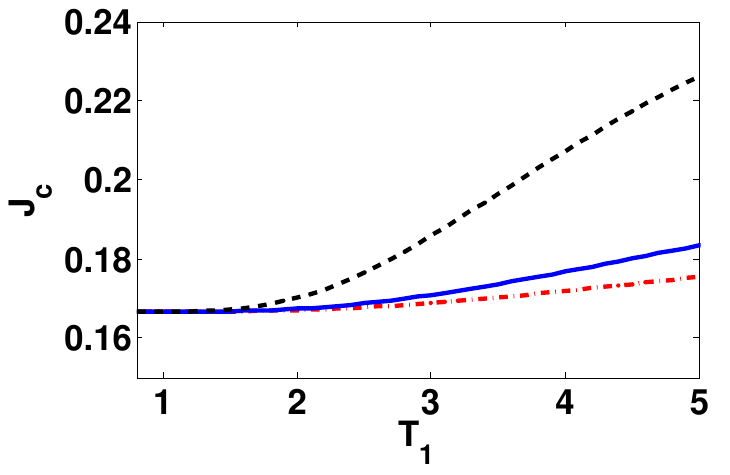}}
		\caption{ The critical $J$ values for which 
		majorisation inequalities hold for the 
		coupled $(1/2,1)$ system, plotted as 
		function of the hot temperature $T_1$. 
		$J_c$ values shown are for the global system  (black, dashed curve)
		and spin-1 (blue curve) and spin-1/2 (red, dot-dashed curve)
        with parameters set as: $B_{1}=5,B_{2}=3$, $T_2 /T_1 = 0.5$.
        At low temperatures, global as well as local systems
        acquire a value $J_c = \Phi/6=0.167$.
        } 
		\label{fig_jmax}
	\end{figure}

	Now, let us analyze the behavior of 
	local work in this range of $J$ values.
	From the analytic expressions as well as
	Fig. (\ref{fig_jmax}), we notice that 
	$J_c^{(1/2)} \approx J_c^{(1)} \approx  J_c \approx  \Phi/6$ at low temperatures. It implies that both local work and the global work are positive in $0\leq J\leq \Phi/6$. For higher temperatures,
	when $J\geq \Phi/6$, 
	 the critical values  of $J$ follow: $J_c^{(1/2)}\leq J_c\leq J_c^{(1)}$, which indicates that the work performed by spin-1/2 can be negative even when the  global work and the 
	 local work due to spin-1 are positive.
	The same conclusion can be justified from the sign of $\cal Z$.
	Thus, it can be proved that
	${\cal Z}\leq 0$ for $J > \Phi/6$ 
	(see Appendix C.2). 
	Applied to Eq. (\ref{eqn105}), this result implies 
	that the local work by spin-1 is {\it always}  positive 
	provided the global work is positive (${\cal X} >0$) for  
	$J > \Phi/6$. 
	Analogously,
	it can be inferred from Eq. (\ref{eqn91})  
	that for values of $J > \Phi/6$, the work performed by spin-1/2 
	{\it can} be negative even when the global work is positive. 
	
	Thus, we can summarize that in the range 
	$0\leq J \leq \Phi/6$,
	both spin-1/2 and spin-1 yield positive work, and so the 
	global work is positive. Beyond this range ($J > \Phi/6$),
	spin-1/2 may yield negative work even when the global work 
	is positive. However, 
	spin-1 yields positive work only if global work is positive. 
	The local work analysis can be extended to the general case
	of coupled $(1/2,s)$ which is described in Appendix C.1.
	
	\section{Enhancement of Otto efficiency}
	\begin{figure}[h!]
		\centering
		{\includegraphics[width=0.7\linewidth]{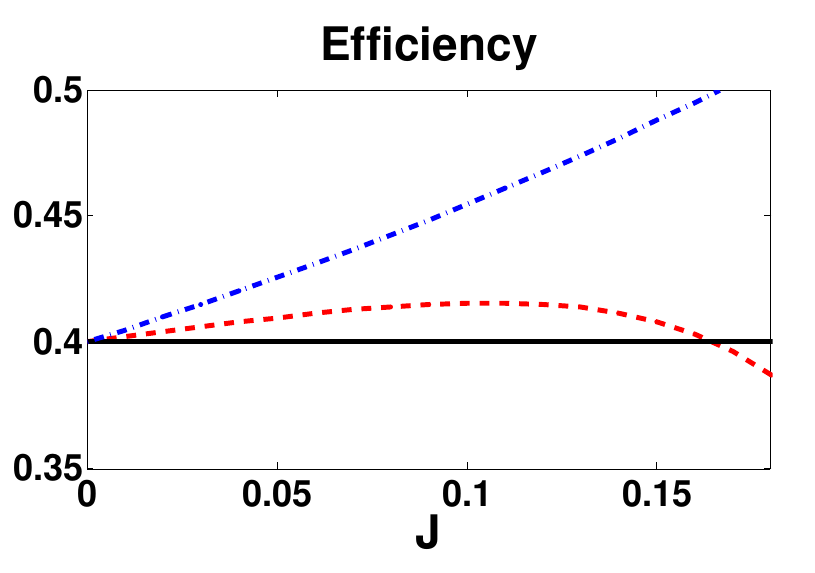}}
		\caption{Quantum Otto efficiency for 
			the coupled $(1/2,1)$ system versus the coupling 
			strength $J$ (red, dashed curve). 
			The parameters are set at $B_{1}=5,B_{2}=3,T_{1}=6,T_{2}=3$. The horizontal line is 
			uncoupled efficiency, $\eta_0 = 1-B_2/B_1$,
			and Carnot bound is equal to 0.5. The 
			upper bound, $\eta_{\rm ub}^{}$, is shown as the blue, 
			dot-dashed curve.} 
		\label{fig_eff}
	\end{figure}
	The quantum feature of exchange coupling 
	as a resource to enhance the Otto efficiency
	has been studied in earlier works.
	In Ref. \cite{PhysRevE.83.031135}, an upper bound for 
	Otto efficiency which is tighter than the 
	Carnot limit was derived. In the following,
	we revisit the feature of efficiency 
	enhancement and provide justification 
	for the upper bound based on the majorisation
	approach.
	
	The efficiency of the QOE, $\eta={|W|}/{Q_{1}}$
	can be expressed using Eqs. (\ref{eqn56}) and (\ref{eqn58}) as:
	%
	\begin{equation} \label{eqn73}
		\eta= {\eta_{0}}
		\left({1-\frac{6J{\cal Y}}{B_{1}{\cal X}}}
		\right)^{-1},
	\end{equation}
	where $\eta_{0}=1-{B_{2}}/{B_{1}}$
	is the efficiency of uncoupled system
	which is the same 
	regardless of the magnitudes of individual spin. 
	We have seen that owing to 
	consistency with the second law, 
	the quantum Otto efficiency is bounded from above by
	the Carnot value, $1-T_2/T_1$. 
	However, heuristics can be applied to this system \cite{2021} 
	to obtain a tighter upper bound on the efficiency,
	which is given by: 
	\begin{equation} \label{eqn79}
		\eta_{\rm ub}^{} = {\eta_{0}^{}}
		\left({1-\frac{6J}{B_{1}}} \right)^{-1} \leq 1-\frac{T_2}{T_1}.
	\end{equation}
	Thus it can be seen that in the range 
	$0\leq J \leq \Phi/6$, the above bound
	is lower than the Carnot value, and 
	is saturated for $J = \Phi/6$, as shown in Fig. \ref{fig_eff}.
	
	Now, we show how the majorisation relation 
	reveals this upper bound for Otto efficiency.
	We have seen ${\cal X}$ is positive  
	in the range $0 \leq J\leq J_c$. 
	The expression for $\eta$ suggests
	that in the presence of coupling ($J>0$),
	the condition ${\cal Y} = (P_{2}-P_{2}^{\prime})+(P_{4}-P_{4}^{\prime})>0$ enhances the efficiency  
	over the uncoupled model ($\eta > \eta_0$). 
	Now, combining $\mathcal{Y} >0$  with global majorisation condition (Eq. (\ref{eqn66})), we can write:
	\begin{equation}
		\frac{P_{2}+P_{4}}{P_{1}}\geq \frac{P_{2}^{\prime}+P_{4}^{\prime}}{P_{1}^{\prime}},
	\end{equation}
	implying that
 	\begin{align*}
		\frac{\exp{\big (\frac{B_{1}+12J}{T_{1}}\big)}+\exp{\big (\frac{-B_{1}+12J}{T_{1}}\big)}}{\exp{\big (\frac{3B_{1}}{T_{1}}\big)}}&\geq\frac{\exp{\big (\frac{B_{2}+12J}{T_{1}}\big)}+\exp{\big (\frac{-B_{2}+12J}{T_{2}}\big)}}{\exp{\big (\frac{3B_{2}}{T_{2}}\big)}},
	\end{align*}
which is satisfied in the range 
$0\leq J\leq J_c$, indicating  
the efficiency enhancement region.
The Otto efficiency is depicted in Fig. \ref{fig_eff}.
	
	Now, consider the expression,   
	$({\cal X} - {\cal Y}) \equiv 	
	3(P_{6}-P_{6}^{\prime})+2(P_{5}-P_{5}^{\prime})+(P_{4}-P_{4}^{\prime})+(P_{3}-P_
	{3}^{\prime})$. Adding the inequalities (\ref{eqn60}), (\ref{eqn61})
	and (\ref{eqn63}), and upon rearranging terms, 
	we obtain ${\cal X} - {\cal Y} \geq 0$, or 
	${\cal X} \geq {\cal Y}$. Then, from Eq. (\ref{eqn73}), 
	this implies that in the domain of the majorisation relation 
	($P \prec P'$),
	the efficiency has the  upper bound given by $\eta_{\rm ub}$.
	The proof can be extended to the general case of $(1/2,s)$
	system, which is discussed in Appendix D.

\section{Conclusions}
QOE is one of the most well studied models of a quantum heat engine. It is based on generalizations of the classical adiabatic and  isochoric processes. Further, a spin-based working medium provides a convenient platform to investigate quantum features and their advantages 
for a QOE.
Thus a QOE based on spin-1/2 particle is well known to have an efficiency, $\eta_0=1-B_2/B_1$ \cite{kieu2004second}. Also, $B_2/T_2>B_1/T_1$ 
is a necessary condition, given that $T_1>T_2$ and $B_1>B_2$. The additional condition guarantees PWC, which can be seen from
the expression for work extracted in a QOE:
\begin{equation}
	\vert W \vert=2(B_1-B_2)\left( \frac{1}{1+e^{2B_1/T_1}}-\frac{1}{1+e^{2B_2/T_2}}\right) \geq 0.
\end{equation} 
We have shown that PWC for a QOE based on an arbitrary spin can be derived from 
the concept of majorisation between $P$ and $P'$, whereby the relation $P\prec P'$ provides a necessary and sufficient condition for PWC. Further, we have expressed the work output in a QOE in terms of the relative entropies $D(P||P')$ and $D(P'||P)$. It is clarified 
in general that $S(P)>S(P')$ is necessary, but not a sufficient condition for PWC. Further, the total entropy generated in a quantum Otto 
cycle is given by the sum total of the two relative entropies. 

Then, we have considered a spin-1/2 interacting with a spin-$s$ via 1-d Heisenberg exchange interaction with isotropic coupling strength $(J>0)$. In this case, majorisation yield PWC, provided we impose $B_2/T_2>B_1/T_1$ and additionally restrict $0\leq J\leq J_c$. The critical value  $J_c$ provides a sufficient range for parameter $J$ such that the majorisation inqualities hold good. We have treated the $s=1$ case in detail and provided expressions for the case with a general $s$ value, in the Appendix.
It is important to remark that in Ref. \cite{2021}, an extreme case scenario was used to infer the permissible range of $J$ as $0\leq J\leq \Phi/6$. Since, $J_c \geq \Phi/6$, so the range inferred in this paper extends the previous range of Ref. \cite{2021}. 

Using the global majorisation inequalities, we have also investigated the local thermodynamics of spins. Thereby, it is possible to infer that spin-1/2 ceases to yield work at a certain $J$ value, while spin-1 continues to output work at larger $J$ values. The global work vanishes in between these two values, giving $J_c^{(1/2)}\leq J_c\leq J_c^{(1)}$.
Again, in previous works \cite{PhysRevE.83.031135, 2021}, an enhancement of efficiency was reported for the coupled model and an upper bound for the Otto efficiency was inferred which is tighter than the Carnot value. In the present work, we have justified this bound using majorisation inequalities. All these results can be extended to a (1/2,$s$) system.

In conclusion, our analysis shows that majorisation relation
can usefully characterize the operational conditions for 
a quantum Otto engine based on spins as the working medium.
The approach based on majorisation is able to highlight key qualitative features of  even the local spin cycles.
 It will be interesting to extend the analysis to other interacting models  \cite{zhang2008entangled, azimi2014quantum, Halpern2019}.  Thus, 
 majorisation is expected to serve as a key heuristic in inferring the thermodynamic features of complex working media in a quantum heat engine. 
	\section*{Acknowledgment}
	SS acknowledges financial support in the form 
	of Senior Research Fellowship from the Council for Scientific and
	Industrial Research (CSIR) via Award No. 09/947(0250)/2020-EMR-I India.
	\appendix
	\label{Appendix}
\section{Work output in terms of relative entropy}
We consider a quasi-static quantum Otto engine (QOE) in the presence 
of two heat reservoirs at temperatures $T_1$ and $T_2 (< T_1)$, as described in the 
main text. 

The heat exchange at hot and cold reservoirs are  given respectively as:
\begin{align} \label{Neqn7}
	Q_{1}&=\sum_{k}\varepsilon_{k}(P_{k}-P_{k}^{\prime}),~~ 
	Q_{2} =\sum_{k}\varepsilon_{k}^{\prime}(P_{k}^{\prime}-P_{k}).
\end{align}
The work output in one quantum Otto cycle is given by:
\begin{equation} \label{fmj1}
	\vert W\vert=\sum_{k}(\varepsilon_{k}-\varepsilon_{k}^{\prime})(P_{k}-P_{k}^{\prime}),
\end{equation}
where
\begin{equation} \label{fmj2}
	P_{k}=\frac{\exp(-\varepsilon_{k}/T_{1})}{Z_{1}}, \quad 
	P_{k}^{\prime}=\frac{\exp(-\varepsilon_{k}^{\prime}/T_{2})}{Z_{2}},
\end{equation}	
are the canonical occupation probabilities for the system while it is in 
equilibrium with hot and cold reservoirs, respectively.
$Z_1$ and $Z_2$ are the corresponding canonical partition sums.
We have set Boltzmann's constant equal to unity. 
The above relations can be inverted as 
\begin{align}
	\varepsilon_{k}=-T_{1}\ln{(P_{k}Z_{1})},& \quad 
	\varepsilon_{k}^{\prime}=-T_{2}\ln{(P_{k}^{\prime}Z_{2})} \label{fmj5},
\end{align}
and substituted in Eq. (\ref{Neqn7}), to obtain  
\begin{align} \label{Nfmj6b}
	Q_1&=\sum_{k}\left (-T_{1}\ln{(P_{k}Z_{1})}\right)(P_{k}-P_{k}^{\prime}) \nonumber \\
	&=\sum_{k} \left( -T_{1}P_{k}\ln P_{k}+T_{1}P_{k}^{\prime
	}\ln P_{k} \right)\nonumber \\
	&=\sum_{k} \left( -T_{1}P_{k}\ln P_{k}+T_{1}P_{k}^{\prime
	}\ln P_{k} +T_{1}P_{k}^{\prime
	}\ln P_{k}^{\prime}-T_{1}P_{k}^{\prime
	}\ln P_{k}^{\prime}\right)
\end{align}

We have applied the normalization conditions $ \sum_{k}P_{k}=\sum_{k}P_{k}^{\prime}=1$, added and subtracted suitable terms to write the above as follows.
\begin{equation}
	Q_1=T_1 (S_1-S_2) - T_1 D(P^{\prime}||P).
\end{equation}
Similarly,
\begin{equation}
	Q_2=-T_2 (S_1-S_2) - T_2D(P||P^{\prime}).
\end{equation}

Substituting Eq.(\ref{fmj5}) in Eq.(\ref{fmj1}) to get work expression:
\begin{align} \label{fmj6b}
	\vert W \vert&=\sum_{k}\left [-T_{1}\ln{(P_{k}Z_{1})}+T_{2}\ln{(P_{k}^{\prime}Z_{2})}\right](P_{k}-P_{k}^{\prime}) \nonumber \\
	&=\sum_{k} \left( -T_{1}P_{k}\ln P_{k}-T_{2}P_{k}^{\prime}\ln P_{k}^{\prime}+T_{1}P_{k}^{\prime
	}\ln P_{k}+T_{2}P_{k}\ln P_{k}^{\prime} \right) \nonumber \\
&=\sum_{k} \left(-T_{1}P_{k}\ln P_{k}-T_{2}P_{k}^{\prime}\ln P_{k}^{\prime}+T_{1}P_{k}^{\prime
}\ln P_{k}+T_{2}P_{k}\ln P_{k}^{\prime}\right)\nonumber\\
& \quad
+\sum_{k}\left (T_{1}P_{k}^{\prime}\ln P_{k}^{\prime}-T_{1}P_{k}^{\prime}\ln P_{k}^{\prime}+T_{2}P_{k}\ln P_{k}-T_{2}P_{k}\ln P_{k}\right).
\end{align}

Upon rearranging the terms on the rhs of the above equation,
we can write
\begin{equation}\label{fmj6}
	\vert W \vert=(T_{1}-T_{2}) (S_{1}-S_{2})-T_{1}D(P^{\prime}||P)-T_{2}D(P||P^{\prime}),
\end{equation}	
where $S_1=-\sum_{k} P_{k} \ln P_{k}$ and $S_2 = -\sum_{k} P_{k}' \ln P_{k}'$
is Shannon entropy of the system 
in contact with hot and cold reservoirs respectively, while
$D(P||P^{\prime})=\sum_k P_{k}
\ln (P_{k}/P_{k}^{\prime})$
and 
$D(P^{\prime}||P)=\sum_k P_{k}^{\prime}
\ln (P_{k}^{\prime}/P_{k})$
are the alternate forms of
Kullback-Leibler divergence (relative entropy) between 
the distributions $P$ and $P'$.

\section{PWC for the coupled $(1/2,s)$ system}
The special case of coupled $(1/2,1)$ system has been described
in the main text.
Here, we considered the more general case 
of two coupled spins, denoted as $(1/2,s)$ system,
whose energy spectrum is shown in Fig. 1.
The total number of energy levels is $n=2(2s+1)$.

	\begin{tikzpicture}
		\draw (0,0)--(4,0);
	\end{tikzpicture}~~$(2s+1)B$
	\\
	
	\begin{tikzpicture}
		\draw (0,0)--(4,0); 
	\end{tikzpicture}~~$(2s-1)B$
	
	\begin{tikzpicture}
		\draw (0,0)--(4,0); 
	\end{tikzpicture}~~$(2s-1)B-4(2s+1)J$\\
	
	\begin{tikzpicture}
		\draw (0,0)--(4,0); 
	\end{tikzpicture}~~$(2s-3)B$
	
	\begin{tikzpicture}
		\draw (0,0)--(4,0); 
	\end{tikzpicture}~~$(2s-3)B-4(2s+1)J$

	~~~~~~~~~~~\textbf{\vdots}
	
	\begin{tikzpicture}
		\draw (0,0)--(4,0); 
	\end{tikzpicture}~~$B$
	
	~~~~~~~~~~~\textbf{\vdots}
	
	\begin{tikzpicture}
		\draw (0,0)--(4,0); 
	\end{tikzpicture}~~$B-4(2s+1)J$\\
	
	\begin{tikzpicture}
		\draw (0,0)--(4,0); 
	\end{tikzpicture}~~$-B$
	
	~~~~~~~~~~~\textbf{\vdots}
	
	\begin{tikzpicture}
		\draw (0,0)--(4,0); 
	\end{tikzpicture}~~$-B-4(2s+1)J$
	
	~~~~~~~~~~~\textbf{\vdots}
	
	\begin{tikzpicture}
		\draw (0,0)--(4,0); 
	\end{tikzpicture}~~$-(2s-3)B$
	
	\begin{tikzpicture}
		\draw (0,0)--(4,0); 
	\end{tikzpicture}~~$-(2s-3)B-4(2s+1)J$\\
	
	\begin{tikzpicture}
		\draw (0,0)--(4,0); 
	\end{tikzpicture}~~$-(2s-1)B$
	
	\begin{tikzpicture}
		\draw (0,0)--(4,0); 
	\end{tikzpicture}~~$-(2s-1)B-4(2s+1)J$\\
	
	\begin{tikzpicture}
		\draw (0,0)--(4,0); 
	\end{tikzpicture}~~$-(2s+1)B$	
	
	\par\noindent
	{Fig. 1: Energy eigenvalue spectrum for coupled $(1/2,s)$ system (see \cite{2021}).}
	
	The heat exchanged at the hot reservoir can be 
	written as
	\begin{equation} \label{feqn9}
		Q_{1}=2B_{1}{\cal X}-4(2s+1)J {\cal Y},
	\end{equation}
	where
	\begin{align}		
		{\cal X} & =\sum_{k=1}^{(n-2)/2} {k}\{(P_{2k}+P_{2k+1})
		-(P_{2k}^{\prime}+P_{2k+1}^{\prime})\}
		+\frac{n}{2} (P_{n} -P_{n}^{\prime}), \label{feqn10} \\
		{\cal Y} &=\sum_{k=1}^{(n-2)/2}(P_{2k}-P_{2k}^{\prime}).
		\label{feqn11}
	\end{align}
	Similarly, the heat exchanged at the cold reservoir is given by $Q_{2}=2B_{2}{\cal X}-4(2s+1)J {\cal Y}$.		
	So, the net work performed in one cycle is given as:
	\begin{equation}
		|W|=2(B_{1}-B_{2}){\cal X}.
		\label{fwglo}
	\end{equation}
	Now, we prove the positive work condition (PWC) i.e. ${\cal X} \geq 0$  for the case $B_{1}>B_{2}$.
	We assume the	
	majorisation relation ($P_{}\prec P_{}^{\prime}$), 
	which implies the validity of the following set of inequalities:
	\begin{align}
		P_{n}&\geq P_{n}^{\prime} \label{feqn16}\\
		P_{n-1}+P_{n}&\geq P_{n}^{\prime}+ P_{n-1}^{\prime} \label{feqn17}\\
		P_{n-2}	+  P_{n-1}+P_{n}&\geq P_{n}^{\prime}+ P_{n-1}^{\prime} + P_{n-2}^{\prime} \label{feqn18}\\
		\quad \vdots \nonumber \\
		P_{2} + P_{3}+ \dots + P_{n-1}+P_{n}&
		\geq P_{n}^{\prime}+ P_{n-1}^{\prime} + \dots + P_{3}^{\prime} + P_{2}^{\prime} . \label{feqn19} 
	\end{align}
	In all, there are $(n-1)$ inequalities
	in the above, which is an odd number 
	since $n=2(2s+1)$ is even. 
	Now, starting from the top,  
	adding up all the alternate inequalities
	(Eqs. (\ref{feqn16}), 
	(\ref{feqn18}), and so on up to (\ref{feqn19})),
	and upon rearranging, we conclude that  ${\cal X} \geq 0$.
	Again, the special case of $s=1$ has been 
	discussed in detail in the main text. 
	
	This proves that the majorisation relation 
	($P_{}\prec P_{}^{\prime}$) implies the PWC for 
	the quantum Otto cycle based on the coupled $(1/2,s)$ system.

	From the normalization condition on probabilities
	and Eq. (\ref{feqn19}), we get 
	
	\begin{equation} \label{fSM1}
		P_1 \leq P_1'.
	\end{equation}
	Combining Eq. (\ref{feqn16}) and Eq. (\ref{fSM1}) we can write
	\begin{equation} \label{fSM2}
		\frac{P_{n}}{P_{1}}\geq \frac{P_{n}^{\prime}}{P_{1}^{\prime}}.
	\end{equation}
	On using the canonical forms of the probabilities in
	the above condition, we obtain
	\begin{equation}\label{fSM3}
		\frac{B_2}{B_1}\geq\frac{T_2}{T_1}
	\end{equation}

	Validity of Eq. (\ref{feqn19})  yield the condition:
	\begin{equation} \label{feqn24}
		0\leq J\leq\frac{\Phi}{2(2s+1)}+\frac{1}{4(2s+1)}\left(\frac{1}{T_2}-\frac{1}{T_1}\right)^{-1}\ln\left(\frac{1+\sum_{m=1}^{s}e^{-(m+1)B_1/T_1}}{1+\sum_{m=1}^{s}e^{-(m+1)B_2/T_2}}\right)\equiv {\cal J}_c,
	\end{equation}
	where $\Phi = (B_2 - B_1 \theta)/(1-\theta)$ and 
	$\theta = T_2/T_1$.
	Thus, for conditions (\ref{fSM3}) and (\ref{feqn24}), 
	all majorisation inequalities hold and give 
	sufficient criteria  for PWC in the case of $(1/2,s)$ system. 
	
	
	\section{Local thermodynamics of individual spins}
	
	\subsection{Local work analysis for (1/2,s) system}
	After showing that the majorisation relation
	for the coupled system implies PWC,
	we proceed to analyze the thermodynamic behavior of
	individual spins. Our interest is 
	to see up to what extent the global 
	majorisation relations determine the 
	operation of each spin. 
	First, we note that 
	the probability distribution for the 
	reduced state of spin-$1/2$ is given by
	\begin{align}
		q_{1} &=\frac{1}{2s+1}\left((2s+1)P_{1}+\sum_{k=1}^{2s}kP_{2k}+\sum_{k=1}^{2s}(2s+1-k)P_{2k+1}\right), \\
		q_2 & = 1 - q_1.
	\end{align}
	Here, $q_1$ is the ground state probability for spin-1/2. 
	Similarly, the probability distribution 
	for the reduced state of spin-$s$ is given by:
	\begin{align}
		r_{1} & =\frac{1}{2s+1}\left[(2s+1)P_{1}+2 s P_{2} + P_{3} \right],
		\label{fgenr1}\\
		r_k  & = \frac{1}{2s+1}\left [ (k-1)P_{2k-2}+(2s)P_{2k-1}+
		(2s-1)P_{2k}+ k P_{2k+1}\right ], 
		\forall k=2,3,\cdots, 2s, \label{fgenrk}\\
		r_{2s+1} &=\frac{1}{2s+1}\left [2 s P_{4s} + P_{4s+1} + 
		(2s+1) P_{4s+2}\right ], \label{fgenrn}
	\end{align}
	where $r_1 > r_2 > \cdots r_{2s} > r_{2s+1}$. 
	Clearly, these expressions reduce to those given in the 
	main text for the case $s=1$.
	
	Then, the work performed by spin-1/2 can be 
	calculated to be:
	\begin{equation} \label{feqn40}
		W_{1/2}^{}=\frac{2}{2s+1} (B_{1}-B_{2})({\cal X}+ \cal Z),
	\end{equation}
	and work by spin-$s$ is given as:
	\begin{equation}  \label{feqn41}
		W_{s}^{}=\frac{2}{2s+1} (B_{1}-B_{2})(2s{\cal X}-\cal Z),
	\end{equation}
	where ${\cal X}$ is given by Eq. (\ref{feqn10}) and 
	
	%
	
	\begin{align}\label{zeqn}
		\cal Z &= 
		\left\{
		\begin{array}{ll}
			& \sum_{k=1}^{n/4-1/2}(n/2-2k)\{(P_{2k}-P_{2k}^{\prime})-(P_{n-2k}-(P_{n-2k}^{
				\prime})\},~~{s=1,2,3..} \\
			& \sum_{k=1}^{n/4-1}(n/2-2k)\{(P_{2k}-P_{2k}^{\prime})-(P_{n-2k}-(P_{n-2k}^{
				\prime})\},~~~{s=1/2,3/2,5/2...}
		\end{array}
		\right.
	\end{align}
	%
	It can be seen that the sum total of the local contributions to work
	add up to yield the global work, $W_{1/2}+W_s = |W|$,
	as given in Eq. (\ref{fwglo}).
	
	\textbf{(i) PWC for spin-1/2}
	
	  Given $B_{1}>B_{2}$,  PWC for spin-1/2 requires,
	\begin{equation} \label{NWfeqn45}
		({\cal X+\cal Z}) \geq 0.
	\end{equation}
	
	Combining global majorisation condition Eq. (\ref{fSM1}) and Eq. (\ref{NWfeqn45}), we obtained sufficient  condition for PWC on coupling constant ($J$),
	\begin{align} \label{Nfeqn24}
		0\leq J\leq\frac{\Phi}{2(2s+1)}+\frac{1}{4(2s+1)}\left(\frac{1}{T_2}-\frac{1}{T_1}\right)^{-1}\ln\left(\frac{2s+1+\sum_{m=1}^{s}(2s-m+1)e^{-(m+2)B_1/T_1}}{2s+1+\sum_{m=1}^{s}(2s-m+1)e^{-(m+2)B_2/T_2}}\right)\nonumber\\
		\equiv {\cal J}_c^{(1/2)}.
	\end{align}
	
	\textbf{(ii) PWC for spin-s} 
	
	Given $B_{1}>B_{2}$,  PWC for spin-s requires,
	\begin{equation} \label{NWfeqn46}
		2s{\cal X-\cal Z}\geq 0.
	\end{equation}
	
	Eq. (\ref{fSM1}) and Eq. (\ref{NWfeqn46}) give sufficient condition on $J$,
	\begin{align} \label{Nfeqn24b}
		0\leq	J\leq\frac{\Phi}{2(2s+1)}+\frac{1}{4(2s+1)}\left(\frac{1}{T_2}-\frac{1}{T_1}\right)^{-1}\ln\left(\frac{1+\sum_{m=1}^{s}(2ms+2m+1)e^{-(m+1)B_1/T_1}}{1+\sum_{m=1}^{s}(2ms+2m+1)e^{-(m+1)B_2/T_2}}\right)\nonumber\\
		\equiv {\cal J}_c^{(1)}.
	\end{align}
	\subsection{Local work analysis for $(1/2,1)$ system}
	In this section, we derive PWC for individual spins, assuming global majorisation relation. Thus we have
	following set of conditions ($P\prec P^{\prime}$),
	\begin{align}
		P_{6}&\geq P_{6}^{\prime} \label{feqn60}\\
		P_{5}+P_{6}&\geq P_{6}^{\prime}+P_{5}^{\prime} \label{feqn61}\\
		P_{4}+P_{5}+P_{6}&\geq P_{6}^{\prime}+P_{5}^{\prime}+P_{4}^{\prime} \label{feqn62}\\
		P_{3}+P_{4}+P_{5}+P_{6}&\geq 
		P_{6}^{\prime}+P_{5}^{\prime}+P_{4}^{\prime}+P_{3}^{\prime} \label{feqn63}\\
		P_{2}+P_{3}+P_{4}+P_{5}+P_{6}&\geq 
		P_{6}^{\prime}+P_{5}^{\prime}+P_{4}^{\prime}+P_{3}^{\prime}+P_{2}^{\prime}. \label{feqn64}
	\end{align}
	Due to normalization of each probability
	distribution, Eq. (\ref{feqn64}) implies:
	\begin{equation} \label{feqn66n}
		P_{1}\leq P_{1}^{\prime}.
	\end{equation}

	\par\noindent
	\textbf{PWC for spin-1/2:}
	
	From Eq. (\ref{feqn40}), the PWC for spin-1/2 requires:
	\begin{align}
		{\cal X} +{\cal Z} &\geq0 \label{feqn92}.
	\end{align}
	On using Eq. (\ref{feqn10}) and Eq. (\ref{zeqn}) for s=1,  we can write
	\begin{equation}
		3P_{6}+2P_{5}+P_{4}+P_{3}+2P_{2}\geq 
		3P_{6}^{\prime}+2P_{5}^{\prime}+P_{4}^{\prime}+P_{3}^{\prime}+2P_{2}^{\prime} \label{feqn94}.
	\end{equation}
	Combining Eq. (\ref{feqn66}) and Eq. (\ref{feqn94}), 
	we get
	\begin{equation} \label{feqn95}
		\frac{	
			3P_{6}+2P_{5}+P_{4}+P_{3}+2P_{2}}{P_{1}}\geq\frac{3P_{6}^{\prime}+2P_{5}^{\prime
			}+P_{4}^{\prime}+P_{3}^{\prime}+2P_{2}^{\prime}}{P_{1}^{\prime}},
	\end{equation}
	which implies the following:
	\begin{align} \label{feqn96}
		\frac{3\exp{\left (\frac{-3B_{1}}{T_{1}}\right)}+2\exp{\left (\frac{-B_{1}}{T_{1}}\right)}+\exp{\left (\frac{-B_{1}+12J}{T_{1}}\right)}+\exp{\left 
				(\frac{B_{1}}{T_{1}}\right)}+2\exp{\left (\frac{B_{1}+12J}{T_{1}}\right
				)}}{\exp{\left 
				(\frac{3B_{1}}{T_{1}}\right
				)}}\nonumber&\geq\\
		\frac{3\exp{\left(\frac{-3B_{2}}{T_{2}}\right)}+2\exp{\left 
				(\frac{-B_{2}}{T_{2}}\right
				)}+\exp{\left(\frac{-B_{2}+12J}{T_{2}}\right)}+\exp{\left 
				(\frac{B_{2}}{T_{2}}\right)}+2\exp{\left (\frac{B_{2}+12J}{T_{2}}\right)}}{\exp{\left 
				(\frac{3B_{2}}{T_{2}}\right)}},
	\end{align}
	%
that can be rearranged as follows:
\begin{align} \label{feqn98}
	&3\left[\exp{\left (\frac{-6B_{1}}{T_{1}}\right)}-\exp{\left 
		(\frac{-6B_{2}}{T_{2}}\right)}\right]+2\left[\exp{\left 
		(\frac{-4B_{1}}{T_{1}}\right)}-\exp{\left (\frac{-4B_{2}}{T_{2}}\right)}\right]+\nonumber\\&
	\left[\exp{\left (\frac{-4B_{1}+12J}{T_{1}}\right)}-\exp{\left 
		(\frac{-4B_{2}+12J}{T_{2}}\right)}\right]+\left[\exp{\left 
		(\frac{-2B_{1}}{T_{1}}\right)}-\exp{\left 
		(\frac{-2B_{2}}{T_{2}}\right)}\right]+\nonumber\\&2\left[\exp{\left 
		(\frac{-2B_{1}+12J}{T_{1}}\right)}-\exp{\left 
		(\frac{-2B_{2}+12J}{T_{2}}\right)}\right]\geq0.
\end{align}
Now, for $B_2/T_2>B_1/T_1$,
the first, second and fourth terms in square brackets above
are positive. Positivity of the third and fifth terms together implies, 
\begin{align} \label{Nfeqn98}
	&\left[\exp{\left (\frac{-4B_{1}+12J}{T_{1}}\right)}-\exp{\left 
		(\frac{-4B_{2}+12J}{T_{2}}\right)}\right]+\nonumber\\&2\left[\exp{\left 
		(\frac{-2B_{1}+12J}{T_{1}}\right)}-\exp{\left 
		(\frac{-2B_{2}+12J}{T_{2}}\right)}\right]\geq0.
\end{align}
Which bound $J$ value as:
\begin{equation} \label{eqnJmx1}
	J\leq\frac{\Phi}{6} +\frac{1}{12}\left(\frac{1}{T_2}-\frac{1}{T_1}\right)^{-1}\ln\left(\frac{2+e^{-2B_1/T_1}}{2+e^{-2B_2/T_2}}\right)\equiv J_c^{(1/2)}.
\end{equation}
This implies that
$0\leq J\leq J_c^{(1/2)}$ is a sufficient condition for 
PWC in case of spin-1/2. 

\par\noindent
\textbf{PWC for spin-1:}

From Eq. (\ref{feqn41}), PWC for spin-1 requires:
\begin{align}
	2{\cal X}-{\cal Z}& \geq 0. 
	\label{feqn107}
\end{align}
On using Eqs. (\ref{feqn10})and (\ref{zeqn}) for s=1, we can write 
\begin{equation}
	6P_{6}+4P_{5}+5P_{4}+2P_{3}+P_{2}\geq 
	6P_{6}^{\prime}+4P_{5}^{\prime}+5P_{4}^{\prime}+2P_{3}^{\prime}+P_{2}^{\prime}. \label{feqn108}
\end{equation}
Combining Eqs. (\ref{feqn66}) and (\ref{feqn108}), we get
\begin{equation} \label{feqn109}	
	\frac{6P_{6}+4P_{5}+5P_{4}+2P_{3}+P_{2}}{P_{1}}\geq\frac{6P_{6}^{\prime}+4P_{5}^
		{\prime}+5P_{4}^{\prime}+2P_{3}^{\prime}+P_{2}^{\prime}}{P_{1}^{\prime}}.
\end{equation}
On rearranging, we get 
\begin{align}
	&6\left[\exp{\left (\frac{-6B_{1}}{T_{1}}\right)}-\exp{\left
		(\frac{-6B_{2}}{T_{2}}\right)}\right]+4\left[\exp{\left 
		(\frac{-4B_{1}}{T_{1}}\right)}-\exp{\left (\frac{-4B_{2}}{T_{2}}\right)}\right]+\nonumber\\& 
	5\left[\exp{\left (\frac{-4B_{1}+12J}{T_{1}}\right)}-\exp{\left 
		(\frac{-4B_{2}+12J}{T_{2}}\right)}\right]+2\left[\exp{\left 
		(\frac{-2B_{1}}{T_{1}}\right)}-\exp{\left 
		(\frac{-2B_{2}}{T_{2}}\right)}\right]+\nonumber\\&\left[\exp{\left 
		(\frac{-2B_{1}+12J}{T_{1}}\right)}-\exp{\left 
		(\frac{-2B_{2}+12J}{T_{2}}\right)}\right]\geq0 \label{feqn112}.
\end{align}
Now, for $B_2/T_2>B_1/T_1$,
the first, second, and fourth terms in 
square brackets above are positive.
Positivity of the third and fifth terms gives,
\begin{align}
	&5\left[\exp{\left (\frac{-4B_{1}+12J}{T_{1}}\right)}-\exp{\left 
		(\frac{-4B_{2}+12J}{T_{2}}\right)}\right]+\nonumber\\&\left[\exp{\left 
		(\frac{-2B_{1}+12J}{T_{1}}\right)}-\exp{\left 
		(\frac{-2B_{2}+12J}{T_{2}}\right)}\right]\geq0 \label{feqn112}.
\end{align}
with the $J$ range given by:
\begin{equation} \label{eqnJmx2}
	0 \leq J\leq\frac{\Phi}{6} +\frac{1}{12}\left(\frac{1}{T_2}-\frac{1}{T_1}\right)^{-1}\ln\left(\frac{1+5e^{-2B_1/T_1}}{1+5e^{-2B_2/T_2}}\right)\equiv J_c^{(1)}.
\end{equation}
This implies that
$0\leq J\leq J_c^{(1)}$ is a sufficient condition for 
PWC in case of the spin-1 subsystem.

In the following, we give explicit proof of 
${\cal Z} \leq 0$ for $J\geq \Phi/6$ in case $(1/2,1)$ system. 
For this system, 
\begin{equation}
	{\cal Z}=(P_2-P_4)-(P_2^{\prime}-P_4^{\prime}). 
	\label{Zfors1}
\end{equation}

\textbf{Proof:}
Let us suppose ${\cal Z} \leq 0$, so that 
\begin{equation*}
	P_{2}-P_{4}\leq P_{2}^{\prime}-P_{4}^{\prime}.
\end{equation*}
Plugging in the explicit forms of the canonical probabilities
in the above, we obtain
\begin{align}	
	\frac{\exp\left(\frac{B_{1}+12J}{T_{1}}\right)-\exp\left(\frac{-B_{1}+12J}{T_{1}}\right)}{\exp\left(\frac{
			3B_{1}}{T_{1}}\right)+\exp\left(\frac{B_{1}+12J}{T_{1}}\right)+\exp\left(\frac{B_{1}}{T_{1}}
		\right)+\exp\left(\frac{-B_{1}+12J}{T_{1}}\right)+\exp\left(\frac{-B_{1}}{T_{1}}\right)+\exp\left(\frac{-3B_{1}}{
			T_{1}}\right)}\nonumber\leq\\	
	\frac{\exp\left(\frac{B_{2}+12J}{T_{2}}\right)-\exp\left(\frac{-B_{2}+12J}{T_{2}}\right)}{\exp\left(\frac{
			3B_{2}}{T_{2}}\right)+\exp\left(\frac{B_{2}+12J}{T_{2}}\right)+\exp\left(\frac{B_{2}}{T_{2}}
		\right)+\exp\left(\frac{-B_{2}+12J}{T_{2}}\right)+\exp\left(\frac{-B_{2}}{T_{2}}\right)+\exp\left(\frac{-3B_{2}}{
			T_{2}}\right)} \label{feqn116}.
\end{align}
Simplify further, we get
\begin{align}	
	\frac{1-\exp\left(\frac{-2B_{1}}{T_{1}}\right)}{\exp(\frac{2B_{1}-12J}{T_{1}})+1+\exp\left(\frac
		{-12J}{T_{1}}\right)+\exp\left(\frac{-2B_{1}}{T_{1}}\right)+\exp\left(\frac{-2B_{1}-12J}{T_{1}}
		\right)+\exp\left(\frac{-4B_{1}-12J}{T_{1}}\right)}\nonumber\leq\\
	\frac{1-\exp\left(\frac{-2B_{2}}{T_{2}}\right)}{\exp\left(\frac{2B_{2}-12J}{T_{2}}\right)+1+\exp\left(\frac
		{-12J}{T_{1}}\right)+\exp\left(\frac{-2B_{2}}{T_{2}}\right)+\exp\left(\frac{-2B_{2}-12J}{T_{2}}
		\right)+\exp\left(\frac{-4B_{2}-12J}{T_{2}}\right)} \label{feqn117}.
\end{align}

The conditions, $T_1 > T_2$, $B_1 > B_2 > 6J$ and ${B_{2}}/{T_{2}}>  {B_{1}}/{T_{1}}$, 
favor the above inequality. However,
the following terms have ambiguous relation:
$$ \exp\left(\frac{2B_{1}-12J}{T_{1}} \right) 
\lessgtr \exp\left(\frac{2B_{2}-12J}{T_{2}}\right).$$
However, if we consider the best case scenario, such that 
these terms also favor the inequality ${\cal Z} \leq 0$, then we 
must have
$$\frac{2B_{1}-12J}{T_{1}} \geq \frac{2B_{2}-12J}{T_{2}},$$
which implies that $J\geq {\Phi}/{6}$.
Thus, for $J\geq {\Phi}/{6}$, we have ${\cal Z} \leq 0$.

The extension of the above proof for the general case ${\cal Z}< 0$ (Eq.(\ref{zeqn})) gives  $J\geq \frac{\Phi}{2(2s+1)}$.

${\cal Z} <0$ implies that the work performed by higher spin will always be  positive when the 
global work is positive (${\cal X}\geq 0$), and the work performed by spin-1/2 can be negative even when the
global work is positive.

\section{An upper bound for Otto efficiency}
The efficiency of QOE is defined as:
$\eta={|W|}/{Q_{1}}$, which can be written as
\begin{equation} \label{feqn26}
	\eta=\frac{(B_{1}-B_{2}){\cal X}}{B_{1}{\cal X}-2(2s+1)J
		{\cal Y}},
\end{equation}
or 
\begin{equation} \label{feqn28}
	\eta={\eta_{0}}
	\left({1-\frac{2(2s+1)J{\cal Y}}{B_{1}{\cal X}}} \right)^{-1},
\end{equation}
where 
$\eta_{0}=1-{B_{2}}/{B_{1}}$ is 
the efficiency of the uncoupled system ($J=0$). ${\cal X}$ and ${\cal Y}$ are defined Eqs. (\ref{feqn10}) and (\ref{feqn11}).
We have seen that
${\cal X} \geq 0$  for 
$0 \leq J\leq {\cal J}_c$ . Therefore, 
${\cal Y}>0$ gives the region where efficiency can be enhanced
over that of the uncoupled system. 
From Eq. (\ref{feqn11}), we note that the expression for 
$\cal Y$ involves only the levels which also depend on parameter
$J$. It is these levels that contribute to a decrease 
in the heat which is not converted into work due
to the fixed nature of parameter $J$. If the net
flow of heat through these levels can be from cold 
to hot, then the efficiency may be enhanced \cite{PhysRevE.83.031135,Oliveira2021}. 

%
From majorisation inequalities and using the expressions
for $\cal X$ and $\cal Y$ given above, we can show
\begin{align}
	{\cal X}&\geq {\cal Y}. \label{feqn37}
\end{align}
The special case of the (1/2,1) system has been described in the paper.
Based on this, we can infer an upper bound for 
the Otto efficiency as 
\begin{equation} \label{feqn38}
	\eta_{\rm ub}^{}={\eta_{0}} \left(
	{1-\frac{2(2s+1)J}{B_{1}}} \right)^{-1} \leq 1-\frac{T_2}{T_1}.
\end{equation} 
The above bound provides a useful upper bound
as long as it stays lower than the Carnot value, i.e.
for $ J \leq \frac{\Phi}{2(2s+1)}$.
	
%


\begin{thebibliography}{83}%
\makeatletter
\providecommand \@ifxundefined [1]{%
 \@ifx{#1\undefined}
}%
\providecommand \@ifnum [1]{%
 \ifnum #1\expandafter \@firstoftwo
 \else \expandafter \@secondoftwo
 \fi
}%
\providecommand \@ifx [1]{%
 \ifx #1\expandafter \@firstoftwo
 \else \expandafter \@secondoftwo
 \fi
}%
\providecommand \natexlab [1]{#1}%
\providecommand \enquote  [1]{``#1''}%
\providecommand \bibnamefont  [1]{#1}%
\providecommand \bibfnamefont [1]{#1}%
\providecommand \citenamefont [1]{#1}%
\providecommand \href@noop [0]{\@secondoftwo}%
\providecommand \href [0]{\begingroup \@sanitize@url \@href}%
\providecommand \@href[1]{\@@startlink{#1}\@@href}%
\providecommand \@@href[1]{\endgroup#1\@@endlink}%
\providecommand \@sanitize@url [0]{\catcode `\\12\catcode `\$12\catcode
  `\&12\catcode `\#12\catcode `\^12\catcode `\_12\catcode `\%12\relax}%
\providecommand \@@startlink[1]{}%
\providecommand \@@endlink[0]{}%
\providecommand \url  [0]{\begingroup\@sanitize@url \@url }%
\providecommand \@url [1]{\endgroup\@href {#1}{\urlprefix }}%
\providecommand \urlprefix  [0]{URL }%
\providecommand \Eprint [0]{\href }%
\providecommand \doibase [0]{https://doi.org/}%
\providecommand \selectlanguage [0]{\@gobble}%
\providecommand \bibinfo  [0]{\@secondoftwo}%
\providecommand \bibfield  [0]{\@secondoftwo}%
\providecommand \translation [1]{[#1]}%
\providecommand \BibitemOpen [0]{}%
\providecommand \bibitemStop [0]{}%
\providecommand \bibitemNoStop [0]{.\EOS\space}%
\providecommand \EOS [0]{\spacefactor3000\relax}%
\providecommand \BibitemShut  [1]{\csname bibitem#1\endcsname}%
\let\auto@bib@innerbib\@empty
\bibitem [{\citenamefont {Vinjanampathy}\ and\ \citenamefont
  {Anders}(2016)}]{Vinjanampathy_2016}%
  \BibitemOpen
  \bibfield  {author} {\bibinfo {author} {\bibfnamefont {S.}~\bibnamefont
  {Vinjanampathy}}\ and\ \bibinfo {author} {\bibfnamefont {J.}~\bibnamefont
  {Anders}},\ }\bibfield  {title} {\bibinfo {title} {Quantum thermodynamics},\
  }\href {https://doi.org/10.1080/00107514.2016.1201896} {\bibfield  {journal}
  {\bibinfo  {journal} {Contemporary Physics}\ }\textbf {\bibinfo {volume}
  {57}},\ \bibinfo {pages} {545} (\bibinfo {year} {2016})}\BibitemShut
  {NoStop}%
\bibitem [{\citenamefont {Kosloff}(2013)}]{Kosloff_2013}%
  \BibitemOpen
  \bibfield  {author} {\bibinfo {author} {\bibfnamefont {R.}~\bibnamefont
  {Kosloff}},\ }\bibfield  {title} {\bibinfo {title} {Quantum thermodynamics: A
  dynamical viewpoint},\ }\href {https://doi.org/10.3390/e15062100} {\bibfield
  {journal} {\bibinfo  {journal} {Entropy}\ }\textbf {\bibinfo {volume} {15}},\
  \bibinfo {pages} {2100} (\bibinfo {year} {2013})}\BibitemShut {NoStop}%
\bibitem [{\citenamefont {Millen}\ and\ \citenamefont
  {Xuereb}(2016)}]{Millen_2016}%
  \BibitemOpen
  \bibfield  {author} {\bibinfo {author} {\bibfnamefont {J.}~\bibnamefont
  {Millen}}\ and\ \bibinfo {author} {\bibfnamefont {A.}~\bibnamefont
  {Xuereb}},\ }\bibfield  {title} {\bibinfo {title} {Perspective on quantum
  thermodynamics},\ }\href {https://doi.org/10.1088/1367-2630/18/1/011002}
  {\bibfield  {journal} {\bibinfo  {journal} {New Journal of Physics}\ }\textbf
  {\bibinfo {volume} {18}},\ \bibinfo {pages} {011002} (\bibinfo {year}
  {2016})}\BibitemShut {NoStop}%
\bibitem [{\citenamefont {Quan}\ \emph {et~al.}(2007)\citenamefont {Quan},
  \citenamefont {Liu}, \citenamefont {Sun},\ and\ \citenamefont
  {Nori}}]{quan2007quantum}%
  \BibitemOpen
  \bibfield  {author} {\bibinfo {author} {\bibfnamefont {H.}~\bibnamefont
  {Quan}}, \bibinfo {author} {\bibfnamefont {Y.-x.}\ \bibnamefont {Liu}},
  \bibinfo {author} {\bibfnamefont {C.}~\bibnamefont {Sun}},\ and\ \bibinfo
  {author} {\bibfnamefont {F.}~\bibnamefont {Nori}},\ }\bibfield  {title}
  {\bibinfo {title} {Quantum thermodynamic cycles and quantum heat engines},\
  }\href@noop {} {\bibfield  {journal} {\bibinfo  {journal} {Phys. Rev. E}\
  }\textbf {\bibinfo {volume} {76}},\ \bibinfo {pages} {031105} (\bibinfo
  {year} {2007})}\BibitemShut {NoStop}%
\bibitem [{\citenamefont {Scovil}\ and\ \citenamefont
  {Schulz-DuBois}(1959)}]{Threelevelmaser}%
  \BibitemOpen
  \bibfield  {author} {\bibinfo {author} {\bibfnamefont {H.~E.~D.}\
  \bibnamefont {Scovil}}\ and\ \bibinfo {author} {\bibfnamefont {E.~O.}\
  \bibnamefont {Schulz-DuBois}},\ }\bibfield  {title} {\bibinfo {title}
  {Three-level masers as heat engines},\ }\href
  {https://doi.org/10.1103/PhysRevLett.2.262} {\bibfield  {journal} {\bibinfo
  {journal} {Phys. Rev. Lett.}\ }\textbf {\bibinfo {volume} {2}},\ \bibinfo
  {pages} {262} (\bibinfo {year} {1959})}\BibitemShut {NoStop}%
\bibitem [{\citenamefont {Alicki}(2014)}]{alicki2014quantum}%
  \BibitemOpen
  \bibfield  {author} {\bibinfo {author} {\bibfnamefont {R.}~\bibnamefont
  {Alicki}},\ }\bibfield  {title} {\bibinfo {title} {Quantum thermodynamics: An
  example of two-level quantum machine},\ }\href@noop {} {\bibfield  {journal}
  {\bibinfo  {journal} {Open Systems \& Information Dynamics}\ }\textbf
  {\bibinfo {volume} {21}},\ \bibinfo {pages} {1440002} (\bibinfo {year}
  {2014})}\BibitemShut {NoStop}%
\bibitem [{\citenamefont {Rivas}(2020)}]{rivas2020strong}%
  \BibitemOpen
  \bibfield  {author} {\bibinfo {author} {\bibfnamefont {{\'A}.}~\bibnamefont
  {Rivas}},\ }\bibfield  {title} {\bibinfo {title} {Strong coupling
  thermodynamics of open quantum systems},\ }\href@noop {} {\bibfield
  {journal} {\bibinfo  {journal} {Phys. Rev. Lett.}\ }\textbf {\bibinfo
  {volume} {124}},\ \bibinfo {pages} {160601} (\bibinfo {year}
  {2020})}\BibitemShut {NoStop}%
\bibitem [{\citenamefont {Goold}\ \emph {et~al.}(2016)\citenamefont {Goold},
  \citenamefont {Huber}, \citenamefont {Riera}, \citenamefont {Del~Rio},\ and\
  \citenamefont {Skrzypczyk}}]{goold2016role}%
  \BibitemOpen
  \bibfield  {author} {\bibinfo {author} {\bibfnamefont {J.}~\bibnamefont
  {Goold}}, \bibinfo {author} {\bibfnamefont {M.}~\bibnamefont {Huber}},
  \bibinfo {author} {\bibfnamefont {A.}~\bibnamefont {Riera}}, \bibinfo
  {author} {\bibfnamefont {L.}~\bibnamefont {Del~Rio}},\ and\ \bibinfo {author}
  {\bibfnamefont {P.}~\bibnamefont {Skrzypczyk}},\ }\bibfield  {title}
  {\bibinfo {title} {The role of quantum information in thermodynamics—a
  topical rev.},\ }\href@noop {} {\bibfield  {journal} {\bibinfo  {journal}
  {Journal of Physics A: Mathematical and Theoretical}\ }\textbf {\bibinfo
  {volume} {49}},\ \bibinfo {pages} {143001} (\bibinfo {year}
  {2016})}\BibitemShut {NoStop}%
\bibitem [{\citenamefont {Jiao}\ \emph {et~al.}(2021)\citenamefont {Jiao},
  \citenamefont {Zhu}, \citenamefont {He}, \citenamefont {Ma},\ and\
  \citenamefont {Wang}}]{PhysRevE.103.032130}%
  \BibitemOpen
  \bibfield  {author} {\bibinfo {author} {\bibfnamefont {G.}~\bibnamefont
  {Jiao}}, \bibinfo {author} {\bibfnamefont {S.}~\bibnamefont {Zhu}}, \bibinfo
  {author} {\bibfnamefont {J.}~\bibnamefont {He}}, \bibinfo {author}
  {\bibfnamefont {Y.}~\bibnamefont {Ma}},\ and\ \bibinfo {author}
  {\bibfnamefont {J.}~\bibnamefont {Wang}},\ }\bibfield  {title} {\bibinfo
  {title} {Fluctuations in irreversible quantum {O}tto engines},\ }\href
  {https://doi.org/10.1103/PhysRevE.103.032130} {\bibfield  {journal} {\bibinfo
   {journal} {Phys. Rev. E}\ }\textbf {\bibinfo {volume} {103}},\ \bibinfo
  {pages} {032130} (\bibinfo {year} {2021})}\BibitemShut {NoStop}%
\bibitem [{\citenamefont {Albash}\ \emph {et~al.}(2012)\citenamefont {Albash},
  \citenamefont {Boixo}, \citenamefont {Lidar},\ and\ \citenamefont
  {Zanardi}}]{Albash_2012}%
  \BibitemOpen
  \bibfield  {author} {\bibinfo {author} {\bibfnamefont {T.}~\bibnamefont
  {Albash}}, \bibinfo {author} {\bibfnamefont {S.}~\bibnamefont {Boixo}},
  \bibinfo {author} {\bibfnamefont {D.~A.}\ \bibnamefont {Lidar}},\ and\
  \bibinfo {author} {\bibfnamefont {P.}~\bibnamefont {Zanardi}},\ }\bibfield
  {title} {\bibinfo {title} {Quantum adiabatic markovian master equations},\
  }\href {https://doi.org/10.1088/1367-2630/14/12/123016} {\bibfield  {journal}
  {\bibinfo  {journal} {New Journal of Physics}\ }\textbf {\bibinfo {volume}
  {14}},\ \bibinfo {pages} {123016} (\bibinfo {year} {2012})}\BibitemShut
  {NoStop}%
\bibitem [{\citenamefont {Chitambar}\ and\ \citenamefont
  {Gour}(2019)}]{RevModPhys.91.025001}%
  \BibitemOpen
  \bibfield  {author} {\bibinfo {author} {\bibfnamefont {E.}~\bibnamefont
  {Chitambar}}\ and\ \bibinfo {author} {\bibfnamefont {G.}~\bibnamefont
  {Gour}},\ }\bibfield  {title} {\bibinfo {title} {Quantum resource theories},\
  }\href {https://doi.org/10.1103/RevModPhys.91.025001} {\bibfield  {journal}
  {\bibinfo  {journal} {Rev. Mod. Phys.}\ }\textbf {\bibinfo {volume} {91}},\
  \bibinfo {pages} {025001} (\bibinfo {year} {2019})}\BibitemShut {NoStop}%
\bibitem [{\citenamefont {Esposito}\ \emph {et~al.}(2010)\citenamefont
  {Esposito}, \citenamefont {Lindenberg},\ and\ \citenamefont {den
  Broeck}}]{Esposito_2010}%
  \BibitemOpen
  \bibfield  {author} {\bibinfo {author} {\bibfnamefont {M.}~\bibnamefont
  {Esposito}}, \bibinfo {author} {\bibfnamefont {K.}~\bibnamefont
  {Lindenberg}},\ and\ \bibinfo {author} {\bibfnamefont {C.~V.}\ \bibnamefont
  {den Broeck}},\ }\bibfield  {title} {\bibinfo {title} {Entropy production as
  correlation between system and reservoir},\ }\href
  {https://doi.org/10.1088/1367-2630/12/1/013013} {\bibfield  {journal}
  {\bibinfo  {journal} {New Journal of Physics}\ }\textbf {\bibinfo {volume}
  {12}},\ \bibinfo {pages} {013013} (\bibinfo {year} {2010})}\BibitemShut
  {NoStop}%
\bibitem [{\citenamefont {Maruyama}\ \emph {et~al.}(2009)\citenamefont
  {Maruyama}, \citenamefont {Nori},\ and\ \citenamefont
  {Vedral}}]{RevModPhys.81.1}%
  \BibitemOpen
  \bibfield  {author} {\bibinfo {author} {\bibfnamefont {K.}~\bibnamefont
  {Maruyama}}, \bibinfo {author} {\bibfnamefont {F.}~\bibnamefont {Nori}},\
  and\ \bibinfo {author} {\bibfnamefont {V.}~\bibnamefont {Vedral}},\
  }\bibfield  {title} {\bibinfo {title} {Colloquium: The physics of maxwell's
  demon and information},\ }\href {https://doi.org/10.1103/RevModPhys.81.1}
  {\bibfield  {journal} {\bibinfo  {journal} {Rev. Mod. Phys.}\ }\textbf
  {\bibinfo {volume} {81}},\ \bibinfo {pages} {1} (\bibinfo {year}
  {2009})}\BibitemShut {NoStop}%
\bibitem [{\citenamefont {Parrondo}\ \emph {et~al.}(2015)\citenamefont
  {Parrondo}, \citenamefont {Horowitz},\ and\ \citenamefont
  {Sagawa}}]{Parrondo}%
  \BibitemOpen
  \bibfield  {author} {\bibinfo {author} {\bibfnamefont {J.~M.}\ \bibnamefont
  {Parrondo}}, \bibinfo {author} {\bibfnamefont {J.}~\bibnamefont {Horowitz}},\
  and\ \bibinfo {author} {\bibfnamefont {T.}~\bibnamefont {Sagawa}},\
  }\bibfield  {title} {\bibinfo {title} {Thermodynamics of information},\
  }\href {https://doi.org/10.1038/nphys3230} {\bibfield  {journal} {\bibinfo
  {journal} {Nature Physics}\ }\textbf {\bibinfo {volume} {11}},\ \bibinfo
  {pages} {131} (\bibinfo {year} {2015})}\BibitemShut {NoStop}%
\bibitem [{\citenamefont {Popescu}\ \emph {et~al.}(2006)\citenamefont
  {Popescu}, \citenamefont {Short},\ and\ \citenamefont
  {Winter}}]{Popescu2006}%
  \BibitemOpen
  \bibfield  {author} {\bibinfo {author} {\bibfnamefont {S.}~\bibnamefont
  {Popescu}}, \bibinfo {author} {\bibfnamefont {A.}~\bibnamefont {Short}},\
  and\ \bibinfo {author} {\bibfnamefont {A.}~\bibnamefont {Winter}},\
  }\bibfield  {title} {\bibinfo {title} {Entanglement and the foundations of
  statistical mechanics},\ }\href {https://doi.org/10.1038/nphys444} {\bibfield
   {journal} {\bibinfo  {journal} {Nature Physics}\ }\textbf {\bibinfo {volume}
  {2}},\ \bibinfo {pages} {754} (\bibinfo {year} {2006})}\BibitemShut {NoStop}%
\bibitem [{\citenamefont {Alicki}(1979)}]{Alicki1979TheQO}%
  \BibitemOpen
  \bibfield  {author} {\bibinfo {author} {\bibfnamefont {R.}~\bibnamefont
  {Alicki}},\ }\bibfield  {title} {\bibinfo {title} {The quantum open system as
  a model of the heat engine},\ }\href@noop {} {\bibfield  {journal} {\bibinfo
  {journal} {Journal of Physics A}\ }\textbf {\bibinfo {volume} {12}} (\bibinfo
  {year} {1979})}\BibitemShut {NoStop}%
\bibitem [{\citenamefont {Allahverdyan}\ \emph {et~al.}(2008)\citenamefont
  {Allahverdyan}, \citenamefont {Johal},\ and\ \citenamefont
  {Mahler}}]{AJM2008}%
  \BibitemOpen
  \bibfield  {author} {\bibinfo {author} {\bibfnamefont {A.~E.}\ \bibnamefont
  {Allahverdyan}}, \bibinfo {author} {\bibfnamefont {R.~S.}\ \bibnamefont
  {Johal}},\ and\ \bibinfo {author} {\bibfnamefont {G.}~\bibnamefont
  {Mahler}},\ }\bibfield  {title} {\bibinfo {title} {Work extremum principle:
  Structure and function of quantum heat engines},\ }\href
  {https://doi.org/10.1103/PhysRevE.77.041118} {\bibfield  {journal} {\bibinfo
  {journal} {Phys. Rev. E}\ }\textbf {\bibinfo {volume} {77}},\ \bibinfo
  {pages} {041118} (\bibinfo {year} {2008})}\BibitemShut {NoStop}%
\bibitem [{\citenamefont {Zhang}\ \emph {et~al.}(2022)\citenamefont {Zhang},
  \citenamefont {Wang}, \citenamefont {Zeng},\ and\ \citenamefont
  {Wang}}]{PRXQuantum.3.030315}%
  \BibitemOpen
  \bibfield  {author} {\bibinfo {author} {\bibfnamefont {K.}~\bibnamefont
  {Zhang}}, \bibinfo {author} {\bibfnamefont {X.}~\bibnamefont {Wang}},
  \bibinfo {author} {\bibfnamefont {Q.}~\bibnamefont {Zeng}},\ and\ \bibinfo
  {author} {\bibfnamefont {J.}~\bibnamefont {Wang}},\ }\bibfield  {title}
  {\bibinfo {title} {Conditional entropy production and quantum fluctuation
  theorem of dissipative information: Theory and experiments},\ }\href
  {https://doi.org/10.1103/PRXQuantum.3.030315} {\bibfield  {journal} {\bibinfo
   {journal} {PRX Quantum}\ }\textbf {\bibinfo {volume} {3}},\ \bibinfo {pages}
  {030315} (\bibinfo {year} {2022})}\BibitemShut {NoStop}%
\bibitem [{\citenamefont {Landi}\ and\ \citenamefont
  {Paternostro}(2021)}]{RevModPhys.93.035008}%
  \BibitemOpen
  \bibfield  {author} {\bibinfo {author} {\bibfnamefont {G.~T.}\ \bibnamefont
  {Landi}}\ and\ \bibinfo {author} {\bibfnamefont {M.}~\bibnamefont
  {Paternostro}},\ }\bibfield  {title} {\bibinfo {title} {Irreversible entropy
  production: From classical to quantum},\ }\href
  {https://doi.org/10.1103/RevModPhys.93.035008} {\bibfield  {journal}
  {\bibinfo  {journal} {Rev. Mod. Phys.}\ }\textbf {\bibinfo {volume} {93}},\
  \bibinfo {pages} {035008} (\bibinfo {year} {2021})}\BibitemShut {NoStop}%
\bibitem [{\citenamefont {Esposito}\ \emph {et~al.}(2015)\citenamefont
  {Esposito}, \citenamefont {Ochoa},\ and\ \citenamefont
  {Galperin}}]{esposito2015quantum}%
  \BibitemOpen
  \bibfield  {author} {\bibinfo {author} {\bibfnamefont {M.}~\bibnamefont
  {Esposito}}, \bibinfo {author} {\bibfnamefont {M.~A.}\ \bibnamefont
  {Ochoa}},\ and\ \bibinfo {author} {\bibfnamefont {M.}~\bibnamefont
  {Galperin}},\ }\bibfield  {title} {\bibinfo {title} {Quantum thermodynamics:
  A nonequilibrium green’s function approach},\ }\href@noop {} {\bibfield
  {journal} {\bibinfo  {journal} {Phys. Rev. Lett.}\ }\textbf {\bibinfo
  {volume} {114}},\ \bibinfo {pages} {080602} (\bibinfo {year}
  {2015})}\BibitemShut {NoStop}%
\bibitem [{\citenamefont {Singh}\ \emph {et~al.}(2020)\citenamefont {Singh},
  \citenamefont {Pandit},\ and\ \citenamefont {Johal}}]{VTJ2020}%
  \BibitemOpen
  \bibfield  {author} {\bibinfo {author} {\bibfnamefont {V.}~\bibnamefont
  {Singh}}, \bibinfo {author} {\bibfnamefont {T.}~\bibnamefont {Pandit}},\ and\
  \bibinfo {author} {\bibfnamefont {R.~S.}\ \bibnamefont {Johal}},\ }\bibfield
  {title} {\bibinfo {title} {Optimal performance of a three-level quantum
  refrigerator},\ }\href {https://doi.org/10.1103/PhysRevE.101.062121}
  {\bibfield  {journal} {\bibinfo  {journal} {Phys. Rev. E}\ }\textbf {\bibinfo
  {volume} {101}},\ \bibinfo {pages} {062121} (\bibinfo {year}
  {2020})}\BibitemShut {NoStop}%
\bibitem [{\citenamefont {Maffei}\ \emph {et~al.}(2021)\citenamefont {Maffei},
  \citenamefont {Camati},\ and\ \citenamefont {Auff\`eves}}]{Maffei2021}%
  \BibitemOpen
  \bibfield  {author} {\bibinfo {author} {\bibfnamefont {M.}~\bibnamefont
  {Maffei}}, \bibinfo {author} {\bibfnamefont {P.~A.}\ \bibnamefont {Camati}},\
  and\ \bibinfo {author} {\bibfnamefont {A.}~\bibnamefont {Auff\`eves}},\
  }\bibfield  {title} {\bibinfo {title} {Probing nonclassical light fields with
  energetic witnesses in waveguide quantum electrodynamics},\ }\href
  {https://doi.org/10.1103/PhysRevResearch.3.L032073} {\bibfield  {journal}
  {\bibinfo  {journal} {Phys. Rev. Research}\ }\textbf {\bibinfo {volume}
  {3}},\ \bibinfo {pages} {L032073} (\bibinfo {year} {2021})}\BibitemShut
  {NoStop}%
\bibitem [{\citenamefont {Rubio}\ \emph {et~al.}(2021)\citenamefont {Rubio},
  \citenamefont {Anders},\ and\ \citenamefont {Correa}}]{Rubio2021}%
  \BibitemOpen
  \bibfield  {author} {\bibinfo {author} {\bibfnamefont {J.}~\bibnamefont
  {Rubio}}, \bibinfo {author} {\bibfnamefont {J.}~\bibnamefont {Anders}},\ and\
  \bibinfo {author} {\bibfnamefont {L.~A.}\ \bibnamefont {Correa}},\ }\bibfield
   {title} {\bibinfo {title} {Global quantum thermometry},\ }\href
  {https://doi.org/10.1103/PhysRevLett.127.190402} {\bibfield  {journal}
  {\bibinfo  {journal} {Phys. Rev. Lett.}\ }\textbf {\bibinfo {volume} {127}},\
  \bibinfo {pages} {190402} (\bibinfo {year} {2021})}\BibitemShut {NoStop}%
\bibitem [{\citenamefont {Alves}\ and\ \citenamefont
  {Landi}(2022)}]{LandiPRA2022}%
  \BibitemOpen
  \bibfield  {author} {\bibinfo {author} {\bibfnamefont {G.~O.}\ \bibnamefont
  {Alves}}\ and\ \bibinfo {author} {\bibfnamefont {G.~T.}\ \bibnamefont
  {Landi}},\ }\bibfield  {title} {\bibinfo {title} {Bayesian estimation for
  collisional thermometry},\ }\href
  {https://doi.org/10.1103/PhysRevA.105.012212} {\bibfield  {journal} {\bibinfo
   {journal} {Phys. Rev. A}\ }\textbf {\bibinfo {volume} {105}},\ \bibinfo
  {pages} {012212} (\bibinfo {year} {2022})}\BibitemShut {NoStop}%
\bibitem [{\citenamefont {Camati}\ \emph {et~al.}(2019)\citenamefont {Camati},
  \citenamefont {Santos},\ and\ \citenamefont {Serra}}]{PhysRevA.99.062103}%
  \BibitemOpen
  \bibfield  {author} {\bibinfo {author} {\bibfnamefont {P.~A.}\ \bibnamefont
  {Camati}}, \bibinfo {author} {\bibfnamefont {J.~F.~G.}\ \bibnamefont
  {Santos}},\ and\ \bibinfo {author} {\bibfnamefont {R.~M.}\ \bibnamefont
  {Serra}},\ }\bibfield  {title} {\bibinfo {title} {Coherence effects in the
  performance of the quantum {O}tto heat engine},\ }\href
  {https://doi.org/10.1103/PhysRevA.99.062103} {\bibfield  {journal} {\bibinfo
  {journal} {Phys. Rev. A}\ }\textbf {\bibinfo {volume} {99}},\ \bibinfo
  {pages} {062103} (\bibinfo {year} {2019})}\BibitemShut {NoStop}%
\bibitem [{\citenamefont {Alet}\ \emph {et~al.}(2021)\citenamefont {Alet},
  \citenamefont {Hanada}, \citenamefont {Jevicki},\ and\ \citenamefont
  {Peng}}]{Alet2021}%
  \BibitemOpen
  \bibfield  {author} {\bibinfo {author} {\bibfnamefont {F.}~\bibnamefont
  {Alet}}, \bibinfo {author} {\bibfnamefont {M.}~\bibnamefont {Hanada}},
  \bibinfo {author} {\bibfnamefont {A.}~\bibnamefont {Jevicki}},\ and\ \bibinfo
  {author} {\bibfnamefont {C.}~\bibnamefont {Peng}},\ }\bibfield  {title}
  {\bibinfo {title} {Entanglement and confinement in coupled quantum systems},\
  }\href {https://doi.org/10.1007/JHEP02(2021)034} {\bibfield  {journal}
  {\bibinfo  {journal} {Journal of High Energy Physics}\ }\textbf {\bibinfo
  {volume} {2021}},\ \bibinfo {pages} {34} (\bibinfo {year}
  {2021})}\BibitemShut {NoStop}%
\bibitem [{\citenamefont {Zhang}\ \emph {et~al.}(2007)\citenamefont {Zhang},
  \citenamefont {Liu}, \citenamefont {Chen},\ and\ \citenamefont
  {Li}}]{zhang2007four}%
  \BibitemOpen
  \bibfield  {author} {\bibinfo {author} {\bibfnamefont {T.}~\bibnamefont
  {Zhang}}, \bibinfo {author} {\bibfnamefont {W.-T.}\ \bibnamefont {Liu}},
  \bibinfo {author} {\bibfnamefont {P.-X.}\ \bibnamefont {Chen}},\ and\
  \bibinfo {author} {\bibfnamefont {C.-Z.}\ \bibnamefont {Li}},\ }\bibfield
  {title} {\bibinfo {title} {Four-level entangled quantum heat engines},\
  }\href@noop {} {\bibfield  {journal} {\bibinfo  {journal} {Phys. Rev. A}\
  }\textbf {\bibinfo {volume} {75}},\ \bibinfo {pages} {062102} (\bibinfo
  {year} {2007})}\BibitemShut {NoStop}%
\bibitem [{\citenamefont {Albayrak}(2013)}]{albayrak2013entangled}%
  \BibitemOpen
  \bibfield  {author} {\bibinfo {author} {\bibfnamefont {E.}~\bibnamefont
  {Albayrak}},\ }\bibfield  {title} {\bibinfo {title} {The entangled quantum
  heat engine in the various heisenberg models for a two-qubit system},\
  }\href@noop {} {\bibfield  {journal} {\bibinfo  {journal} {International
  Journal of Quantum Information}\ }\textbf {\bibinfo {volume} {11}},\ \bibinfo
  {pages} {1350021} (\bibinfo {year} {2013})}\BibitemShut {NoStop}%
\bibitem [{\citenamefont {de~la Cruz}\ and\ \citenamefont
  {Martin-Delgado}(2014)}]{Diaz_de_la_Cruz_2014}%
  \BibitemOpen
  \bibfield  {author} {\bibinfo {author} {\bibfnamefont {J.~M.~D.}\
  \bibnamefont {de~la Cruz}}\ and\ \bibinfo {author} {\bibfnamefont {M.~A.}\
  \bibnamefont {Martin-Delgado}},\ }\bibfield  {title} {\bibinfo {title}
  {Quantum-information engines with many-body states attaining optimal
  extractable work with quantum control},\ }\bibfield  {journal} {\bibinfo
  {journal} {Physical Review A}\ }\textbf {\bibinfo {volume} {89}},\ \href
  {https://doi.org/10.1103/physreva.89.032327} {10.1103/physreva.89.032327}
  (\bibinfo {year} {2014})\BibitemShut {NoStop}%
\bibitem [{\citenamefont {Campisi}\ \emph {et~al.}(2015)\citenamefont
  {Campisi}, \citenamefont {Pekola},\ and\ \citenamefont {Fazio}}]{Pekola2015}%
  \BibitemOpen
  \bibfield  {author} {\bibinfo {author} {\bibfnamefont {M.}~\bibnamefont
  {Campisi}}, \bibinfo {author} {\bibfnamefont {J.}~\bibnamefont {Pekola}},\
  and\ \bibinfo {author} {\bibfnamefont {R.}~\bibnamefont {Fazio}},\ }\bibfield
   {title} {\bibinfo {title} {Nonequilibrium fluctuations in quantum heat
  engines: Theory, example, and possible solid state experiments},\ }\href
  {https://doi.org/10.1088/1367-2630/17/3/035012} {\bibfield  {journal}
  {\bibinfo  {journal} {New Journal of Physics}\ }\textbf {\bibinfo {volume}
  {17}} (\bibinfo {year} {2015})}\BibitemShut {NoStop}%
\bibitem [{\citenamefont {Myers}\ and\ \citenamefont
  {Deffner}(2020)}]{Myers_BOF2020}%
  \BibitemOpen
  \bibfield  {author} {\bibinfo {author} {\bibfnamefont {N.~M.}\ \bibnamefont
  {Myers}}\ and\ \bibinfo {author} {\bibfnamefont {S.}~\bibnamefont
  {Deffner}},\ }\bibfield  {title} {\bibinfo {title} {Bosons outperform
  fermions: The thermodynamic advantage of symmetry},\ }\href
  {https://doi.org/10.1103/PhysRevE.101.012110} {\bibfield  {journal} {\bibinfo
   {journal} {Phys. Rev. E}\ }\textbf {\bibinfo {volume} {101}},\ \bibinfo
  {pages} {012110} (\bibinfo {year} {2020})}\BibitemShut {NoStop}%
\bibitem [{\citenamefont {Watanabe}\ \emph {et~al.}(2020)\citenamefont
  {Watanabe}, \citenamefont {Venkatesh}, \citenamefont {Talkner}, \citenamefont
  {Hwang},\ and\ \citenamefont {del Campo}}]{WatanabePRL2020}%
  \BibitemOpen
  \bibfield  {author} {\bibinfo {author} {\bibfnamefont {G.}~\bibnamefont
  {Watanabe}}, \bibinfo {author} {\bibfnamefont {B.~P.}\ \bibnamefont
  {Venkatesh}}, \bibinfo {author} {\bibfnamefont {P.}~\bibnamefont {Talkner}},
  \bibinfo {author} {\bibfnamefont {M.-J.}\ \bibnamefont {Hwang}},\ and\
  \bibinfo {author} {\bibfnamefont {A.}~\bibnamefont {del Campo}},\ }\bibfield
  {title} {\bibinfo {title} {Quantum statistical enhancement of the collective
  performance of multiple bosonic engines},\ }\href
  {https://doi.org/10.1103/PhysRevLett.124.210603} {\bibfield  {journal}
  {\bibinfo  {journal} {Phys. Rev. Lett.}\ }\textbf {\bibinfo {volume} {124}},\
  \bibinfo {pages} {210603} (\bibinfo {year} {2020})}\BibitemShut {NoStop}%
\bibitem [{\citenamefont {Wu}\ \emph {et~al.}(2014)\citenamefont {Wu},
  \citenamefont {He}, \citenamefont {Ma},\ and\ \citenamefont
  {Wang}}]{PhysRevE.90.062134}%
  \BibitemOpen
  \bibfield  {author} {\bibinfo {author} {\bibfnamefont {F.}~\bibnamefont
  {Wu}}, \bibinfo {author} {\bibfnamefont {J.}~\bibnamefont {He}}, \bibinfo
  {author} {\bibfnamefont {Y.}~\bibnamefont {Ma}},\ and\ \bibinfo {author}
  {\bibfnamefont {J.}~\bibnamefont {Wang}},\ }\bibfield  {title} {\bibinfo
  {title} {Efficiency at maximum power of a quantum {O}tto cycle within
  finite-time or irreversible thermodynamics},\ }\href
  {https://doi.org/10.1103/PhysRevE.90.062134} {\bibfield  {journal} {\bibinfo
  {journal} {Phys. Rev. E}\ }\textbf {\bibinfo {volume} {90}},\ \bibinfo
  {pages} {062134} (\bibinfo {year} {2014})}\BibitemShut {NoStop}%
\bibitem [{\citenamefont {Pe\~na}\ \emph {et~al.}(2020)\citenamefont {Pe\~na},
  \citenamefont {Zambrano}, \citenamefont {Negrete}, \citenamefont {De~Chiara},
  \citenamefont {Orellana},\ and\ \citenamefont
  {Vargas}}]{PhysRevE.101.012116}%
  \BibitemOpen
  \bibfield  {author} {\bibinfo {author} {\bibfnamefont {F.~J.}\ \bibnamefont
  {Pe\~na}}, \bibinfo {author} {\bibfnamefont {D.}~\bibnamefont {Zambrano}},
  \bibinfo {author} {\bibfnamefont {O.}~\bibnamefont {Negrete}}, \bibinfo
  {author} {\bibfnamefont {G.}~\bibnamefont {De~Chiara}}, \bibinfo {author}
  {\bibfnamefont {P.~A.}\ \bibnamefont {Orellana}},\ and\ \bibinfo {author}
  {\bibfnamefont {P.}~\bibnamefont {Vargas}},\ }\bibfield  {title} {\bibinfo
  {title} {Quasistatic and quantum-adiabatic {O}tto engine for a
  two-dimensional material: The case of a graphene quantum dot},\ }\href
  {https://doi.org/10.1103/PhysRevE.101.012116} {\bibfield  {journal} {\bibinfo
   {journal} {Phys. Rev. E}\ }\textbf {\bibinfo {volume} {101}},\ \bibinfo
  {pages} {012116} (\bibinfo {year} {2020})}\BibitemShut {NoStop}%
\bibitem [{\citenamefont {Das}\ and\ \citenamefont
  {Mukherjee}(2020)}]{das2020quantum}%
  \BibitemOpen
  \bibfield  {author} {\bibinfo {author} {\bibfnamefont {A.}~\bibnamefont
  {Das}}\ and\ \bibinfo {author} {\bibfnamefont {V.}~\bibnamefont
  {Mukherjee}},\ }\bibfield  {title} {\bibinfo {title} {Quantum-enhanced
  finite-time {O}tto cycle},\ }\href@noop {} {\bibfield  {journal} {\bibinfo
  {journal} {Phys. Rev. Research}\ }\textbf {\bibinfo {volume} {2}},\ \bibinfo
  {pages} {033083} (\bibinfo {year} {2020})}\BibitemShut {NoStop}%
\bibitem [{\citenamefont {Chand}\ \emph {et~al.}(2021)\citenamefont {Chand},
  \citenamefont {Dasgupta},\ and\ \citenamefont {Biswas}}]{chand2021finite}%
  \BibitemOpen
  \bibfield  {author} {\bibinfo {author} {\bibfnamefont {S.}~\bibnamefont
  {Chand}}, \bibinfo {author} {\bibfnamefont {S.}~\bibnamefont {Dasgupta}},\
  and\ \bibinfo {author} {\bibfnamefont {A.}~\bibnamefont {Biswas}},\
  }\bibfield  {title} {\bibinfo {title} {Finite-time performance of a
  single-ion quantum {O}tto engine},\ }\href@noop {} {\bibfield  {journal}
  {\bibinfo  {journal} {Phys. Rev. E}\ }\textbf {\bibinfo {volume} {103}},\
  \bibinfo {pages} {032144} (\bibinfo {year} {2021})}\BibitemShut {NoStop}%
\bibitem [{\citenamefont {Lee}\ \emph {et~al.}(2020)\citenamefont {Lee},
  \citenamefont {Ha}, \citenamefont {Park},\ and\ \citenamefont
  {Jeong}}]{lee2020finite}%
  \BibitemOpen
  \bibfield  {author} {\bibinfo {author} {\bibfnamefont {S.}~\bibnamefont
  {Lee}}, \bibinfo {author} {\bibfnamefont {M.}~\bibnamefont {Ha}}, \bibinfo
  {author} {\bibfnamefont {J.-M.}\ \bibnamefont {Park}},\ and\ \bibinfo
  {author} {\bibfnamefont {H.}~\bibnamefont {Jeong}},\ }\bibfield  {title}
  {\bibinfo {title} {Finite-time quantum {O}tto engine: Surpassing the
  quasistatic efficiency due to friction},\ }\href@noop {} {\bibfield
  {journal} {\bibinfo  {journal} {Phys. Rev. E}\ }\textbf {\bibinfo {volume}
  {101}},\ \bibinfo {pages} {022127} (\bibinfo {year} {2020})}\BibitemShut
  {NoStop}%
\bibitem [{\citenamefont {T{\"u}rkpen{\c{c}}e}\ and\ \citenamefont
  {Altintas}(2019)}]{turkpencce2019coupled}%
  \BibitemOpen
  \bibfield  {author} {\bibinfo {author} {\bibfnamefont {D.}~\bibnamefont
  {T{\"u}rkpen{\c{c}}e}}\ and\ \bibinfo {author} {\bibfnamefont
  {F.}~\bibnamefont {Altintas}},\ }\bibfield  {title} {\bibinfo {title}
  {Coupled quantum {O}tto heat engine and refrigerator with inner friction},\
  }\href@noop {} {\bibfield  {journal} {\bibinfo  {journal} {Quantum
  Information Processing}\ }\textbf {\bibinfo {volume} {18}},\ \bibinfo {pages}
  {255} (\bibinfo {year} {2019})}\BibitemShut {NoStop}%
\bibitem [{\citenamefont {Thomas}\ and\ \citenamefont
  {Johal}(2014)}]{thomas2014friction}%
  \BibitemOpen
  \bibfield  {author} {\bibinfo {author} {\bibfnamefont {G.}~\bibnamefont
  {Thomas}}\ and\ \bibinfo {author} {\bibfnamefont {R.~S.}\ \bibnamefont
  {Johal}},\ }\bibfield  {title} {\bibinfo {title} {Friction due to
  inhomogeneous driving of coupled spins in a quantum heat engine},\
  }\href@noop {} {\bibfield  {journal} {\bibinfo  {journal} {The European Phys.
  Journal B}\ }\textbf {\bibinfo {volume} {87}},\ \bibinfo {pages} {166}
  (\bibinfo {year} {2014})}\BibitemShut {NoStop}%
\bibitem [{\citenamefont {Geva}\ and\ \citenamefont
  {Kosloff}(1992)}]{geva1992quantum}%
  \BibitemOpen
  \bibfield  {author} {\bibinfo {author} {\bibfnamefont {E.}~\bibnamefont
  {Geva}}\ and\ \bibinfo {author} {\bibfnamefont {R.}~\bibnamefont {Kosloff}},\
  }\bibfield  {title} {\bibinfo {title} {A quantum-mechanical heat engine
  operating in finite time. a model consisting of spin-1/2 systems as the
  working fluid},\ }\href@noop {} {\bibfield  {journal} {\bibinfo  {journal}
  {The Journal of {C}hemical {P}hysics}\ }\textbf {\bibinfo {volume} {96}},\
  \bibinfo {pages} {3054} (\bibinfo {year} {1992})}\BibitemShut {NoStop}%
\bibitem [{\citenamefont {Feldmann}\ and\ \citenamefont
  {Kosloff}(2000)}]{feldmann2000performance}%
  \BibitemOpen
  \bibfield  {author} {\bibinfo {author} {\bibfnamefont {T.}~\bibnamefont
  {Feldmann}}\ and\ \bibinfo {author} {\bibfnamefont {R.}~\bibnamefont
  {Kosloff}},\ }\bibfield  {title} {\bibinfo {title} {Performance of discrete
  heat engines and heat pumps in finite time},\ }\href@noop {} {\bibfield
  {journal} {\bibinfo  {journal} {Phys. Rev. E}\ }\textbf {\bibinfo {volume}
  {61}},\ \bibinfo {pages} {4774} (\bibinfo {year} {2000})}\BibitemShut
  {NoStop}%
\bibitem [{\citenamefont {{\c{C}}akmak}\ \emph {et~al.}(2017)\citenamefont
  {{\c{C}}akmak}, \citenamefont {Altintas}, \citenamefont {Gen{\c{c}}ten},\
  and\ \citenamefont {M{\"u}stecapl{\i}o{\u{g}}lu}}]{ccakmak2017irreversible}%
  \BibitemOpen
  \bibfield  {author} {\bibinfo {author} {\bibfnamefont {S.}~\bibnamefont
  {{\c{C}}akmak}}, \bibinfo {author} {\bibfnamefont {F.}~\bibnamefont
  {Altintas}}, \bibinfo {author} {\bibfnamefont {A.}~\bibnamefont
  {Gen{\c{c}}ten}},\ and\ \bibinfo {author} {\bibfnamefont {{\"O}.~E.}\
  \bibnamefont {M{\"u}stecapl{\i}o{\u{g}}lu}},\ }\bibfield  {title} {\bibinfo
  {title} {Irreversible work and internal friction in a quantum {O}tto cycle of
  a single arbitrary spin},\ }\href@noop {} {\bibfield  {journal} {\bibinfo
  {journal} {The European Phys. Journal D}\ }\textbf {\bibinfo {volume} {71}},\
  \bibinfo {pages} {75} (\bibinfo {year} {2017})}\BibitemShut {NoStop}%
\bibitem [{\citenamefont {Shastri}\ and\ \citenamefont
  {Venkatesh}(2022)}]{Shastri2022}%
  \BibitemOpen
  \bibfield  {author} {\bibinfo {author} {\bibfnamefont {R.}~\bibnamefont
  {Shastri}}\ and\ \bibinfo {author} {\bibfnamefont {B.~P.}\ \bibnamefont
  {Venkatesh}},\ }\bibfield  {title} {\bibinfo {title} {Optimization of
  asymmetric quantum otto engine cycles},\ }\href
  {https://doi.org/10.1103/PhysRevE.106.024123} {\bibfield  {journal} {\bibinfo
   {journal} {Phys. Rev. E}\ }\textbf {\bibinfo {volume} {106}},\ \bibinfo
  {pages} {024123} (\bibinfo {year} {2022})}\BibitemShut {NoStop}%
\bibitem [{\citenamefont {Solfanelli}\ \emph {et~al.}(2020)\citenamefont
  {Solfanelli}, \citenamefont {Falsetti},\ and\ \citenamefont
  {Campisi}}]{PhysRevB.101.054513}%
  \BibitemOpen
  \bibfield  {author} {\bibinfo {author} {\bibfnamefont {A.}~\bibnamefont
  {Solfanelli}}, \bibinfo {author} {\bibfnamefont {M.}~\bibnamefont
  {Falsetti}},\ and\ \bibinfo {author} {\bibfnamefont {M.}~\bibnamefont
  {Campisi}},\ }\bibfield  {title} {\bibinfo {title} {Nonadiabatic single-qubit
  quantum otto engine},\ }\href {https://doi.org/10.1103/PhysRevB.101.054513}
  {\bibfield  {journal} {\bibinfo  {journal} {Phys. Rev. B}\ }\textbf {\bibinfo
  {volume} {101}},\ \bibinfo {pages} {054513} (\bibinfo {year}
  {2020})}\BibitemShut {NoStop}%
\bibitem [{\citenamefont {Papadatos}(2021)}]{Papadatos_2021}%
  \BibitemOpen
  \bibfield  {author} {\bibinfo {author} {\bibfnamefont {N.}~\bibnamefont
  {Papadatos}},\ }\bibfield  {title} {\bibinfo {title} {The quantum otto heat
  engine with a relativistically moving thermal bath},\ }\href
  {https://doi.org/10.1007/s10773-021-04969-9} {\bibfield  {journal} {\bibinfo
  {journal} {International Journal of Theoretical Physics}\ }\textbf {\bibinfo
  {volume} {60}},\ \bibinfo {pages} {4210} (\bibinfo {year}
  {2021})}\BibitemShut {NoStop}%
\bibitem [{\citenamefont {Das}\ and\ \citenamefont {Ghosh}(2019)}]{e21111131}%
  \BibitemOpen
  \bibfield  {author} {\bibinfo {author} {\bibfnamefont {A.}~\bibnamefont
  {Das}}\ and\ \bibinfo {author} {\bibfnamefont {S.}~\bibnamefont {Ghosh}},\
  }\bibfield  {title} {\bibinfo {title} {Measurement based quantum heat engine
  with coupled working medium},\ }\bibfield  {journal} {\bibinfo  {journal}
  {Entropy}\ }\textbf {\bibinfo {volume} {21}},\ \href
  {https://doi.org/10.3390/e21111131} {10.3390/e21111131} (\bibinfo {year}
  {2019})\BibitemShut {NoStop}%
\bibitem [{\citenamefont {Peña}\ \emph {et~al.}(2020)\citenamefont {Peña},
  \citenamefont {Negrete}, \citenamefont {Cortés},\ and\ \citenamefont
  {Vargas}}]{e22070755}%
  \BibitemOpen
  \bibfield  {author} {\bibinfo {author} {\bibfnamefont {F.~J.}\ \bibnamefont
  {Peña}}, \bibinfo {author} {\bibfnamefont {O.}~\bibnamefont {Negrete}},
  \bibinfo {author} {\bibfnamefont {N.}~\bibnamefont {Cortés}},\ and\ \bibinfo
  {author} {\bibfnamefont {P.}~\bibnamefont {Vargas}},\ }\bibfield  {title}
  {\bibinfo {title} {{O}tto engine: Classical and quantum approach},\
  }\bibfield  {journal} {\bibinfo  {journal} {Entropy}\ }\textbf {\bibinfo
  {volume} {22}},\ \href {https://doi.org/10.3390/e22070755}
  {10.3390/e22070755} (\bibinfo {year} {2020})\BibitemShut {NoStop}%
\bibitem [{\citenamefont {Lin}\ and\ \citenamefont
  {Chen}(2003)}]{lin2003performance}%
  \BibitemOpen
  \bibfield  {author} {\bibinfo {author} {\bibfnamefont {B.}~\bibnamefont
  {Lin}}\ and\ \bibinfo {author} {\bibfnamefont {J.}~\bibnamefont {Chen}},\
  }\bibfield  {title} {\bibinfo {title} {Performance analysis of an
  irreversible quantum heat engine working with harmonic oscillators},\
  }\href@noop {} {\bibfield  {journal} {\bibinfo  {journal} {Phys. Rev. E}\
  }\textbf {\bibinfo {volume} {67}},\ \bibinfo {pages} {046105} (\bibinfo
  {year} {2003})}\BibitemShut {NoStop}%
\bibitem [{\citenamefont {Rezek}\ and\ \citenamefont
  {Kosloff}(2006)}]{Rezek2006}%
  \BibitemOpen
  \bibfield  {author} {\bibinfo {author} {\bibfnamefont {Y.}~\bibnamefont
  {Rezek}}\ and\ \bibinfo {author} {\bibfnamefont {R.}~\bibnamefont
  {Kosloff}},\ }\bibfield  {title} {\bibinfo {title} {Irreversible performance
  of a quantum harmonic heat engine},\ }\href
  {https://doi.org/10.1088/1367-2630/8/5/083} {\bibfield  {journal} {\bibinfo
  {journal} {New Journal of Physics}\ }\textbf {\bibinfo {volume} {8}},\
  \bibinfo {pages} {83} (\bibinfo {year} {2006})}\BibitemShut {NoStop}%
\bibitem [{\citenamefont {Zhang}(2008)}]{zhang2008entangled}%
  \BibitemOpen
  \bibfield  {author} {\bibinfo {author} {\bibfnamefont {G.-F.}\ \bibnamefont
  {Zhang}},\ }\bibfield  {title} {\bibinfo {title} {Entangled quantum heat
  engines based on two two-spin systems with {Z}yaloshinski-{M}oriya
  anisotropic antisymmetric interaction},\ }\href@noop {} {\bibfield  {journal}
  {\bibinfo  {journal} {The European Phys. Journal D}\ }\textbf {\bibinfo
  {volume} {49}},\ \bibinfo {pages} {123} (\bibinfo {year} {2008})}\BibitemShut
  {NoStop}%
\bibitem [{\citenamefont {H{\"u}bner}\ \emph {et~al.}(2014)\citenamefont
  {H{\"u}bner}, \citenamefont {Lefkidis}, \citenamefont {Dong}, \citenamefont
  {Chaudhuri}, \citenamefont {Chotorlishvili},\ and\ \citenamefont
  {Berakdar}}]{hubner2014spin}%
  \BibitemOpen
  \bibfield  {author} {\bibinfo {author} {\bibfnamefont {W.}~\bibnamefont
  {H{\"u}bner}}, \bibinfo {author} {\bibfnamefont {G.}~\bibnamefont
  {Lefkidis}}, \bibinfo {author} {\bibfnamefont {C.}~\bibnamefont {Dong}},
  \bibinfo {author} {\bibfnamefont {D.}~\bibnamefont {Chaudhuri}}, \bibinfo
  {author} {\bibfnamefont {L.}~\bibnamefont {Chotorlishvili}},\ and\ \bibinfo
  {author} {\bibfnamefont {J.}~\bibnamefont {Berakdar}},\ }\bibfield  {title}
  {\bibinfo {title} {Spin-dependent {O}tto quantum heat engine based on a
  molecular substance},\ }\href@noop {} {\bibfield  {journal} {\bibinfo
  {journal} {Phys. Rev. B}\ }\textbf {\bibinfo {volume} {90}},\ \bibinfo
  {pages} {024401} (\bibinfo {year} {2014})}\BibitemShut {NoStop}%
\bibitem [{\citenamefont {Azimi}\ \emph {et~al.}(2014)\citenamefont {Azimi},
  \citenamefont {Chotorlishvili}, \citenamefont {Mishra}, \citenamefont
  {Vekua}, \citenamefont {H{\"u}bner},\ and\ \citenamefont
  {Berakdar}}]{azimi2014quantum}%
  \BibitemOpen
  \bibfield  {author} {\bibinfo {author} {\bibfnamefont {M.}~\bibnamefont
  {Azimi}}, \bibinfo {author} {\bibfnamefont {L.}~\bibnamefont
  {Chotorlishvili}}, \bibinfo {author} {\bibfnamefont {S.~K.}\ \bibnamefont
  {Mishra}}, \bibinfo {author} {\bibfnamefont {T.}~\bibnamefont {Vekua}},
  \bibinfo {author} {\bibfnamefont {W.}~\bibnamefont {H{\"u}bner}},\ and\
  \bibinfo {author} {\bibfnamefont {J.}~\bibnamefont {Berakdar}},\ }\bibfield
  {title} {\bibinfo {title} {Quantum {O}tto heat engine based on a multiferroic
  chain working substance},\ }\href@noop {} {\bibfield  {journal} {\bibinfo
  {journal} {New Journal of Physics}\ }\textbf {\bibinfo {volume} {16}},\
  \bibinfo {pages} {063018} (\bibinfo {year} {2014})}\BibitemShut {NoStop}%
\bibitem [{\citenamefont {Insinga}\ \emph {et~al.}(2016)\citenamefont
  {Insinga}, \citenamefont {Andresen},\ and\ \citenamefont
  {Salamon}}]{insinga2016thermodynamical}%
  \BibitemOpen
  \bibfield  {author} {\bibinfo {author} {\bibfnamefont {A.}~\bibnamefont
  {Insinga}}, \bibinfo {author} {\bibfnamefont {B.}~\bibnamefont {Andresen}},\
  and\ \bibinfo {author} {\bibfnamefont {P.}~\bibnamefont {Salamon}},\
  }\bibfield  {title} {\bibinfo {title} {Thermodynamical analysis of a quantum
  heat engine based on harmonic oscillators},\ }\href@noop {} {\bibfield
  {journal} {\bibinfo  {journal} {Phys. Rev. E}\ }\textbf {\bibinfo {volume}
  {94}},\ \bibinfo {pages} {012119} (\bibinfo {year} {2016})}\BibitemShut
  {NoStop}%
\bibitem [{\citenamefont {Mehta}\ and\ \citenamefont
  {Johal}(2017)}]{mehta2017quantum}%
  \BibitemOpen
  \bibfield  {author} {\bibinfo {author} {\bibfnamefont {V.}~\bibnamefont
  {Mehta}}\ and\ \bibinfo {author} {\bibfnamefont {R.~S.}\ \bibnamefont
  {Johal}},\ }\bibfield  {title} {\bibinfo {title} {Quantum {O}tto engine with
  exchange coupling in the presence of level degeneracy},\ }\href@noop {}
  {\bibfield  {journal} {\bibinfo  {journal} {Phys. Rev. E}\ }\textbf {\bibinfo
  {volume} {96}},\ \bibinfo {pages} {032110} (\bibinfo {year}
  {2017})}\BibitemShut {NoStop}%
\bibitem [{\citenamefont {Peterson}\ \emph {et~al.}(2019)\citenamefont
  {Peterson}, \citenamefont {Batalh\~ao}, \citenamefont {Herrera},
  \citenamefont {Souza}, \citenamefont {Sarthour}, \citenamefont {Oliveira},\
  and\ \citenamefont {Serra}}]{peterson2019experimental}%
  \BibitemOpen
  \bibfield  {author} {\bibinfo {author} {\bibfnamefont {J.~P.~S.}\
  \bibnamefont {Peterson}}, \bibinfo {author} {\bibfnamefont {T.~B.}\
  \bibnamefont {Batalh\~ao}}, \bibinfo {author} {\bibfnamefont
  {M.}~\bibnamefont {Herrera}}, \bibinfo {author} {\bibfnamefont {A.~M.}\
  \bibnamefont {Souza}}, \bibinfo {author} {\bibfnamefont {R.~S.}\ \bibnamefont
  {Sarthour}}, \bibinfo {author} {\bibfnamefont {I.~S.}\ \bibnamefont
  {Oliveira}},\ and\ \bibinfo {author} {\bibfnamefont {R.~M.}\ \bibnamefont
  {Serra}},\ }\bibfield  {title} {\bibinfo {title} {Experimental
  characterization of a spin quantum heat engine},\ }\href
  {https://doi.org/10.1103/PhysRevLett.123.240601} {\bibfield  {journal}
  {\bibinfo  {journal} {Phys. Rev. Lett.}\ }\textbf {\bibinfo {volume} {123}},\
  \bibinfo {pages} {240601} (\bibinfo {year} {2019})}\BibitemShut {NoStop}%
\bibitem [{\citenamefont {Myers}\ \emph {et~al.}(2022)\citenamefont {Myers},
  \citenamefont {Abah},\ and\ \citenamefont {Deffner}}]{MyersAVS2022}%
  \BibitemOpen
  \bibfield  {author} {\bibinfo {author} {\bibfnamefont {N.~M.}\ \bibnamefont
  {Myers}}, \bibinfo {author} {\bibfnamefont {O.}~\bibnamefont {Abah}},\ and\
  \bibinfo {author} {\bibfnamefont {S.}~\bibnamefont {Deffner}},\ }\bibfield
  {title} {\bibinfo {title} {Quantum thermodynamic devices: From theoretical
  proposals to experimental reality},\ }\href
  {https://doi.org/10.1116/5.0083192} {\bibfield  {journal} {\bibinfo
  {journal} {AVS Quantum Science}\ }\textbf {\bibinfo {volume} {4}},\ \bibinfo
  {pages} {027101} (\bibinfo {year} {2022})},\ \Eprint
  {https://arxiv.org/abs/https://doi.org/10.1116/5.0083192}
  {https://doi.org/10.1116/5.0083192} \BibitemShut {NoStop}%
\bibitem [{\citenamefont {Marshall}\ \emph {et~al.}(2011)\citenamefont
  {Marshall}, \citenamefont {Olkin},\ and\ \citenamefont
  {Arnold}}]{Marshallmajorisationbook}%
  \BibitemOpen
  \bibfield  {author} {\bibinfo {author} {\bibfnamefont {A.~W.}\ \bibnamefont
  {Marshall}}, \bibinfo {author} {\bibfnamefont {I.}~\bibnamefont {Olkin}},\
  and\ \bibinfo {author} {\bibfnamefont {B.~C.}\ \bibnamefont {Arnold}},\
  }\bibfield  {title} {\bibinfo {title} {Inequalities: Theory of majorisation
  and its applications}\ }(\bibinfo  {publisher} {Springer Series in
  Statistics, Springer, New York},\ \bibinfo {year} {2011})\BibitemShut
  {NoStop}%
\bibitem [{\citenamefont {Sagawa}(2020)}]{TSagava}%
  \BibitemOpen
  \bibfield  {author} {\bibinfo {author} {\bibfnamefont {T.}~\bibnamefont
  {Sagawa}},\ }\bibfield  {title} {\bibinfo {title} {Entropy, divergence, and
  majorisation in classical and quantum thermodynamics}\ }(\bibinfo
  {publisher} {SpringerBriefs in Mathematical Physics, Springer Singapore},\
  \bibinfo {year} {2020})\BibitemShut {NoStop}%
\bibitem [{\citenamefont {Bhatia}(1996)}]{Bhatia1996MatrixA}%
  \BibitemOpen
  \bibfield  {author} {\bibinfo {author} {\bibfnamefont {R.}~\bibnamefont
  {Bhatia}},\ }\bibfield  {title} {\bibinfo {title} {Matrix analysis}\
  }(\bibinfo  {publisher} {Springer New York, NY},\ \bibinfo {year}
  {1996})\BibitemShut {NoStop}%
\bibitem [{\citenamefont {Buscemi}\ and\ \citenamefont
  {Gour}(2017)}]{PhysRevA.95.012110}%
  \BibitemOpen
  \bibfield  {author} {\bibinfo {author} {\bibfnamefont {F.}~\bibnamefont
  {Buscemi}}\ and\ \bibinfo {author} {\bibfnamefont {G.}~\bibnamefont {Gour}},\
  }\bibfield  {title} {\bibinfo {title} {Quantum relative lorenz curves},\
  }\href {https://doi.org/10.1103/PhysRevA.95.012110} {\bibfield  {journal}
  {\bibinfo  {journal} {Phys. Rev. A}\ }\textbf {\bibinfo {volume} {95}},\
  \bibinfo {pages} {012110} (\bibinfo {year} {2017})}\BibitemShut {NoStop}%
\bibitem [{\citenamefont {Joe}(1990)}]{Joe1990MajorisationAD}%
  \BibitemOpen
  \bibfield  {author} {\bibinfo {author} {\bibfnamefont {H.}~\bibnamefont
  {Joe}},\ }\bibfield  {title} {\bibinfo {title} {Majorisation and
  divergence},\ }\href@noop {} {\bibfield  {journal} {\bibinfo  {journal}
  {Journal of Mathematical Analysis and Applications}\ }\textbf {\bibinfo
  {volume} {148}},\ \bibinfo {pages} {287} (\bibinfo {year}
  {1990})}\BibitemShut {NoStop}%
\bibitem [{\citenamefont {Shiraishi}(2020)}]{2020}%
  \BibitemOpen
  \bibfield  {author} {\bibinfo {author} {\bibfnamefont {N.}~\bibnamefont
  {Shiraishi}},\ }\bibfield  {title} {\bibinfo {title} {Two constructive proofs
  on d-majorisation and thermo-majorisation},\ }\href
  {https://doi.org/10.1088/1751-8121/abb041} {\bibfield  {journal} {\bibinfo
  {journal} {Journal of Physics A: Mathematical and Theoretical}\ }\textbf
  {\bibinfo {volume} {53}},\ \bibinfo {pages} {425301} (\bibinfo {year}
  {2020})}\BibitemShut {NoStop}%
\bibitem [{\citenamefont {Egloff}\ \emph {et~al.}(2015)\citenamefont {Egloff},
  \citenamefont {Dahlsten}, \citenamefont {Renner},\ and\ \citenamefont
  {Vedral}}]{vedral2015}%
  \BibitemOpen
  \bibfield  {author} {\bibinfo {author} {\bibfnamefont {D.}~\bibnamefont
  {Egloff}}, \bibinfo {author} {\bibfnamefont {O.~C.~O.}\ \bibnamefont
  {Dahlsten}}, \bibinfo {author} {\bibfnamefont {R.}~\bibnamefont {Renner}},\
  and\ \bibinfo {author} {\bibfnamefont {V.}~\bibnamefont {Vedral}},\
  }\bibfield  {title} {\bibinfo {title} {A measure of majorisation emerging
  from single-shot statistical mechanics},\ }\href
  {https://doi.org/10.1088/1367-2630/17/7/073001} {\bibfield  {journal}
  {\bibinfo  {journal} {New Journal of Physics}\ }\textbf {\bibinfo {volume}
  {17}},\ \bibinfo {pages} {073001} (\bibinfo {year} {2015})}\BibitemShut
  {NoStop}%
\bibitem [{\citenamefont {Nielsen}\ and\ \citenamefont
  {Vidal}(2001)}]{10.5555/2011326.2011331}%
  \BibitemOpen
  \bibfield  {author} {\bibinfo {author} {\bibfnamefont {M.~A.}\ \bibnamefont
  {Nielsen}}\ and\ \bibinfo {author} {\bibfnamefont {G.}~\bibnamefont
  {Vidal}},\ }\bibfield  {title} {\bibinfo {title} {Majorisation and the
  interconversion of bipartite states},\ }\href@noop {} {\bibfield  {journal}
  {\bibinfo  {journal} {Quantum Info. Comput.}\ }\textbf {\bibinfo {volume}
  {1}},\ \bibinfo {pages} {76–93} (\bibinfo {year} {2001})}\BibitemShut
  {NoStop}%
\bibitem [{\citenamefont {Rethinasamy}\ and\ \citenamefont
  {Wilde}(2020)}]{PhysRevResearch.2.033455}%
  \BibitemOpen
  \bibfield  {author} {\bibinfo {author} {\bibfnamefont {S.}~\bibnamefont
  {Rethinasamy}}\ and\ \bibinfo {author} {\bibfnamefont {M.~M.}\ \bibnamefont
  {Wilde}},\ }\bibfield  {title} {\bibinfo {title} {Relative entropy and
  catalytic relative majorisation},\ }\href
  {https://doi.org/10.1103/PhysRevResearch.2.033455} {\bibfield  {journal}
  {\bibinfo  {journal} {Phys. Rev. Research}\ }\textbf {\bibinfo {volume}
  {2}},\ \bibinfo {pages} {033455} (\bibinfo {year} {2020})}\BibitemShut
  {NoStop}%
\bibitem [{\citenamefont {Renes}(2016)}]{2016}%
  \BibitemOpen
  \bibfield  {author} {\bibinfo {author} {\bibfnamefont {J.~M.}\ \bibnamefont
  {Renes}},\ }\bibfield  {title} {\bibinfo {title} {Relative submajorisation
  and its use in quantum resource theories},\ }\href
  {https://doi.org/10.1063/1.4972295} {\bibfield  {journal} {\bibinfo
  {journal} {Journal of Mathematical Physics}\ }\textbf {\bibinfo {volume}
  {57}},\ \bibinfo {pages} {122202} (\bibinfo {year} {2016})}\BibitemShut
  {NoStop}%
\bibitem [{\citenamefont {Ruch}\ \emph {et~al.}(1980)\citenamefont {Ruch},
  \citenamefont {Schranner},\ and\ \citenamefont {Seligman}}]{RUCH1980222}%
  \BibitemOpen
  \bibfield  {author} {\bibinfo {author} {\bibfnamefont {E.}~\bibnamefont
  {Ruch}}, \bibinfo {author} {\bibfnamefont {R.}~\bibnamefont {Schranner}},\
  and\ \bibinfo {author} {\bibfnamefont {T.~H.}\ \bibnamefont {Seligman}},\
  }\bibfield  {title} {\bibinfo {title} {Generalization of a theorem by hardy,
  littlewood, and pólya},\ }\href
  {https://doi.org/10.1016/0022-247X(80)90075-X} {\bibfield  {journal}
  {\bibinfo  {journal} {Journal of Mathematical Analysis and Applications}\
  }\textbf {\bibinfo {volume} {76}},\ \bibinfo {pages} {222–229} (\bibinfo
  {year} {1980})}\BibitemShut {NoStop}%
\bibitem [{\citenamefont {Nielsen}(1999)}]{PhysRevLett.83.436}%
  \BibitemOpen
  \bibfield  {author} {\bibinfo {author} {\bibfnamefont {M.~A.}\ \bibnamefont
  {Nielsen}},\ }\bibfield  {title} {\bibinfo {title} {Conditions for a class of
  entanglement transformations},\ }\href
  {https://doi.org/10.1103/PhysRevLett.83.436} {\bibfield  {journal} {\bibinfo
  {journal} {Phys. Rev. Lett.}\ }\textbf {\bibinfo {volume} {83}},\ \bibinfo
  {pages} {436} (\bibinfo {year} {1999})}\BibitemShut {NoStop}%
\bibitem [{\citenamefont {Du}\ \emph {et~al.}(2015)\citenamefont {Du},
  \citenamefont {Bai},\ and\ \citenamefont {Guo}}]{PhysRevA.91.052120}%
  \BibitemOpen
  \bibfield  {author} {\bibinfo {author} {\bibfnamefont {S.}~\bibnamefont
  {Du}}, \bibinfo {author} {\bibfnamefont {Z.}~\bibnamefont {Bai}},\ and\
  \bibinfo {author} {\bibfnamefont {Y.}~\bibnamefont {Guo}},\ }\bibfield
  {title} {\bibinfo {title} {Conditions for coherence transformations under
  incoherent operations},\ }\href {https://doi.org/10.1103/PhysRevA.91.052120}
  {\bibfield  {journal} {\bibinfo  {journal} {Phys. Rev. A}\ }\textbf {\bibinfo
  {volume} {91}},\ \bibinfo {pages} {052120} (\bibinfo {year}
  {2015})}\BibitemShut {NoStop}%
\bibitem [{\citenamefont {Jonathan}\ and\ \citenamefont
  {Plenio}(1999)}]{PhysRevLett.83.3566}%
  \BibitemOpen
  \bibfield  {author} {\bibinfo {author} {\bibfnamefont {D.}~\bibnamefont
  {Jonathan}}\ and\ \bibinfo {author} {\bibfnamefont {M.~B.}\ \bibnamefont
  {Plenio}},\ }\bibfield  {title} {\bibinfo {title} {Entanglement-assisted
  local manipulation of pure quantum states},\ }\href
  {https://doi.org/10.1103/PhysRevLett.83.3566} {\bibfield  {journal} {\bibinfo
   {journal} {Phys. Rev. Lett.}\ }\textbf {\bibinfo {volume} {83}},\ \bibinfo
  {pages} {3566} (\bibinfo {year} {1999})}\BibitemShut {NoStop}%
\bibitem [{\citenamefont {Horodecki}\ \emph {et~al.}(2003)\citenamefont
  {Horodecki}, \citenamefont {Horodecki},\ and\ \citenamefont
  {Oppenheim}}]{Horodecki2003}%
  \BibitemOpen
  \bibfield  {author} {\bibinfo {author} {\bibfnamefont {M.}~\bibnamefont
  {Horodecki}}, \bibinfo {author} {\bibfnamefont {P.}~\bibnamefont
  {Horodecki}},\ and\ \bibinfo {author} {\bibfnamefont {J.}~\bibnamefont
  {Oppenheim}},\ }\bibfield  {title} {\bibinfo {title} {Reversible
  transformations from pure to mixed states and the unique measure of
  information},\ }\href {https://doi.org/10.1103/PhysRevA.67.062104} {\bibfield
   {journal} {\bibinfo  {journal} {Phys. Rev. A}\ }\textbf {\bibinfo {volume}
  {67}},\ \bibinfo {pages} {062104} (\bibinfo {year} {2003})}\BibitemShut
  {NoStop}%
\bibitem [{\citenamefont {Gour}\ \emph {et~al.}(2018)\citenamefont {Gour},
  \citenamefont {Jennings}, \citenamefont {Buscemi}, \citenamefont {Duan},\
  and\ \citenamefont {Marvian}}]{majorisationcomplete}%
  \BibitemOpen
  \bibfield  {author} {\bibinfo {author} {\bibfnamefont {G.}~\bibnamefont
  {Gour}}, \bibinfo {author} {\bibfnamefont {D.}~\bibnamefont {Jennings}},
  \bibinfo {author} {\bibfnamefont {F.}~\bibnamefont {Buscemi}}, \bibinfo
  {author} {\bibfnamefont {R.}~\bibnamefont {Duan}},\ and\ \bibinfo {author}
  {\bibfnamefont {I.}~\bibnamefont {Marvian}},\ }\bibfield  {title} {\bibinfo
  {title} {Quantum majorisation and a complete set of entropic conditions for
  quantum thermodynamics},\ }\bibfield  {journal} {\bibinfo  {journal} {Nature
  Communications}\ }\textbf {\bibinfo {volume} {9}},\ \href
  {https://doi.org/10.1038/s41467-018-06261-7} {10.1038/s41467-018-06261-7}
  (\bibinfo {year} {2018})\BibitemShut {NoStop}%
\bibitem [{\citenamefont {Brandão}\ \emph {et~al.}(2015)\citenamefont
  {Brandão}, \citenamefont {Horodecki}, \citenamefont {Ng}, \citenamefont
  {Oppenheim},\ and\ \citenamefont {Wehner}}]{2015}%
  \BibitemOpen
  \bibfield  {author} {\bibinfo {author} {\bibfnamefont {F.}~\bibnamefont
  {Brandão}}, \bibinfo {author} {\bibfnamefont {M.}~\bibnamefont {Horodecki}},
  \bibinfo {author} {\bibfnamefont {N.}~\bibnamefont {Ng}}, \bibinfo {author}
  {\bibfnamefont {J.}~\bibnamefont {Oppenheim}},\ and\ \bibinfo {author}
  {\bibfnamefont {S.}~\bibnamefont {Wehner}},\ }\bibfield  {title} {\bibinfo
  {title} {The second laws of quantum thermodynamics},\ }\href
  {https://doi.org/10.1073/pnas.1411728112} {\bibfield  {journal} {\bibinfo
  {journal} {Proceedings of the National Academy of Sciences}\ }\textbf
  {\bibinfo {volume} {112}},\ \bibinfo {pages} {3275–3279} (\bibinfo {year}
  {2015})}\BibitemShut {NoStop}%
\bibitem [{\citenamefont {Horodecki}\ and\ \citenamefont
  {Oppenheim}(2013)}]{2013}%
  \BibitemOpen
  \bibfield  {author} {\bibinfo {author} {\bibfnamefont {M.}~\bibnamefont
  {Horodecki}}\ and\ \bibinfo {author} {\bibfnamefont {J.}~\bibnamefont
  {Oppenheim}},\ }\bibfield  {title} {\bibinfo {title} {Fundamental limitations
  for quantum and nanoscale thermodynamics},\ }\bibfield  {journal} {\bibinfo
  {journal} {Nature Communications}\ }\textbf {\bibinfo {volume} {4}},\ \href
  {https://doi.org/10.1038/ncomms3059} {10.1038/ncomms3059} (\bibinfo {year}
  {2013})\BibitemShut {NoStop}%
\bibitem [{\citenamefont {Zemansky}(1968)}]{Zemansky}%
  \BibitemOpen
  \bibfield  {author} {\bibinfo {author} {\bibfnamefont {M.~W.}\ \bibnamefont
  {Zemansky}},\ }\bibfield  {title} {\bibinfo {title} {Heat and thermodynamics;
  an intermediate textbook}\ }(\bibinfo  {publisher} {McGraw-Hill New York},\
  \bibinfo {year} {1968})\BibitemShut {NoStop}%
\bibitem [{\citenamefont {{Born}}\ and\ \citenamefont
  {{Fock}}(1928)}]{Adiabatic_Fock}%
  \BibitemOpen
  \bibfield  {author} {\bibinfo {author} {\bibfnamefont {M.}~\bibnamefont
  {{Born}}}\ and\ \bibinfo {author} {\bibfnamefont {V.}~\bibnamefont
  {{Fock}}},\ }\bibfield  {title} {\bibinfo {title} {{Beweis des
  Adiabatensatzes}},\ }\href {https://doi.org/10.1007/BF01343193} {\bibfield
  {journal} {\bibinfo  {journal} {Zeitschrift fur Physik}\ }\textbf {\bibinfo
  {volume} {51}},\ \bibinfo {pages} {165} (\bibinfo {year} {1928})}\BibitemShut
  {NoStop}%
\bibitem [{\citenamefont {Kullback}\ and\ \citenamefont
  {Leibler}(1951)}]{10.1214/aoms/1177729694}%
  \BibitemOpen
  \bibfield  {author} {\bibinfo {author} {\bibfnamefont {S.}~\bibnamefont
  {Kullback}}\ and\ \bibinfo {author} {\bibfnamefont {R.~A.}\ \bibnamefont
  {Leibler}},\ }\bibfield  {title} {\bibinfo {title} {{On Information and
  Sufficiency}},\ }\href {https://doi.org/10.1214/aoms/1177729694} {\bibfield
  {journal} {\bibinfo  {journal} {The Annals of Mathematical Statistics}\
  }\textbf {\bibinfo {volume} {22}},\ \bibinfo {pages} {79 – 86} (\bibinfo
  {year} {1951})}\BibitemShut {NoStop}%
\bibitem [{\citenamefont {Kieu}(2004)}]{kieu2004second}%
  \BibitemOpen
  \bibfield  {author} {\bibinfo {author} {\bibfnamefont {T.~D.}\ \bibnamefont
  {Kieu}},\ }\bibfield  {title} {\bibinfo {title} {The second law, {M}axwell's
  demon, and work derivable from quantum heat engines},\ }\href@noop {}
  {\bibfield  {journal} {\bibinfo  {journal} {Phys. Rev. Lett.}\ }\textbf
  {\bibinfo {volume} {93}},\ \bibinfo {pages} {140403} (\bibinfo {year}
  {2004})}\BibitemShut {NoStop}%
\bibitem [{\citenamefont {Johal}\ and\ \citenamefont {Mehta}(2021)}]{2021}%
  \BibitemOpen
  \bibfield  {author} {\bibinfo {author} {\bibfnamefont {R.~S.}\ \bibnamefont
  {Johal}}\ and\ \bibinfo {author} {\bibfnamefont {V.}~\bibnamefont {Mehta}},\
  }\bibfield  {title} {\bibinfo {title} {Quantum heat engines with complex
  working media, complete {O}tto cycles and heuristics},\ }\href
  {https://doi.org/10.3390/e23091149} {\bibfield  {journal} {\bibinfo
  {journal} {Entropy}\ }\textbf {\bibinfo {volume} {23}},\ \bibinfo {pages}
  {1149} (\bibinfo {year} {2021})}\BibitemShut {NoStop}%
\bibitem [{\citenamefont {Thomas}\ and\ \citenamefont
  {Johal}(2011)}]{PhysRevE.83.031135}%
  \BibitemOpen
  \bibfield  {author} {\bibinfo {author} {\bibfnamefont {G.}~\bibnamefont
  {Thomas}}\ and\ \bibinfo {author} {\bibfnamefont {R.~S.}\ \bibnamefont
  {Johal}},\ }\bibfield  {title} {\bibinfo {title} {Coupled quantum {O}tto
  cycle},\ }\href {https://doi.org/10.1103/PhysRevE.83.031135} {\bibfield
  {journal} {\bibinfo  {journal} {Phys. Rev. E}\ }\textbf {\bibinfo {volume}
  {83}},\ \bibinfo {pages} {031135} (\bibinfo {year} {2011})}\BibitemShut
  {NoStop}%
\bibitem [{\citenamefont {Altintas}\ and\ \citenamefont {M\"ustecapl\ifmmode
  \imath \else \i \fi{}o\ifmmode~\breve{g}\else
  \u{g}\fi{}lu}(2015)}]{PhysRevE.92.022142}%
  \BibitemOpen
  \bibfield  {author} {\bibinfo {author} {\bibfnamefont {F.}~\bibnamefont
  {Altintas}}\ and\ \bibinfo {author} {\bibfnamefont {O.~E.}\ \bibnamefont
  {M\"ustecapl\ifmmode \imath \else \i \fi{}o\ifmmode~\breve{g}\else
  \u{g}\fi{}lu}},\ }\bibfield  {title} {\bibinfo {title} {General formalism of
  local thermodynamics with an example: Quantum {O}tto engine with a spin-$1/2$
  coupled to an arbitrary spin},\ }\href
  {https://doi.org/10.1103/PhysRevE.92.022142} {\bibfield  {journal} {\bibinfo
  {journal} {Phys. Rev. E}\ }\textbf {\bibinfo {volume} {92}},\ \bibinfo
  {pages} {022142} (\bibinfo {year} {2015})}\BibitemShut {NoStop}%
\bibitem [{\citenamefont {Yunger~Halpern}\ \emph {et~al.}(2019)\citenamefont
  {Yunger~Halpern}, \citenamefont {White}, \citenamefont {Gopalakrishnan},\
  and\ \citenamefont {Refael}}]{Halpern2019}%
  \BibitemOpen
  \bibfield  {author} {\bibinfo {author} {\bibfnamefont {N.}~\bibnamefont
  {Yunger~Halpern}}, \bibinfo {author} {\bibfnamefont {C.~D.}\ \bibnamefont
  {White}}, \bibinfo {author} {\bibfnamefont {S.}~\bibnamefont
  {Gopalakrishnan}},\ and\ \bibinfo {author} {\bibfnamefont {G.}~\bibnamefont
  {Refael}},\ }\bibfield  {title} {\bibinfo {title} {Quantum engine based on
  many-body localization},\ }\href {https://doi.org/10.1103/PhysRevB.99.024203}
  {\bibfield  {journal} {\bibinfo  {journal} {Phys. Rev. B}\ }\textbf {\bibinfo
  {volume} {99}},\ \bibinfo {pages} {024203} (\bibinfo {year}
  {2019})}\BibitemShut {NoStop}%
\bibitem [{\citenamefont {de~Oliveira}\ and\ \citenamefont
  {Jonathan}(2021)}]{Oliveira2021}%
  \BibitemOpen
  \bibfield  {author} {\bibinfo {author} {\bibfnamefont {T.~R.}\ \bibnamefont
  {de~Oliveira}}\ and\ \bibinfo {author} {\bibfnamefont {D.}~\bibnamefont
  {Jonathan}},\ }\bibfield  {title} {\bibinfo {title} {Efficiency gain and
  bidirectional operation of quantum engines with decoupled internal levels},\
  }\href {https://doi.org/10.1103/PhysRevE.104.044133} {\bibfield  {journal}
  {\bibinfo  {journal} {Phys. Rev. E}\ }\textbf {\bibinfo {volume} {104}},\
  \bibinfo {pages} {044133} (\bibinfo {year} {2021})}\BibitemShut {NoStop}%
\end{thebibliography}


\end{document}